\documentclass[fleqn,usenatbib]{mnras}

\usepackage{newtxtext,newtxmath}
\usepackage[T1]{fontenc}

\DeclareRobustCommand{\VAN}[3]{#2}
\let\VANthebibliography\thebibliography
\def\thebibliography{\DeclareRobustCommand{\VAN}[3]{##3}\VANthebibliography}

\usepackage{graphicx}
\usepackage{amsmath}
\usepackage{subcaption}
\usepackage{cleveref}
\usepackage{multirow}
\usepackage{lscape}
\usepackage{rotating}

\newcommand{\mJypbm}{\text{mJy}\,\text{beam}$^{-1}$}

\newcommand{\fig}{Fig.}

\newcommand{\sect}{Section}
\newcommand{\sects}{Sections}

\title{Low radio frequency detections of known pulsars and identification of new candidates with GLEAM-X: GP}
\author[S. Mantovanini et al.]{
Silvia Mantovanini,$^{1}$\thanks{E-mail: silvia.mantovanini@postgrad.curtin.edu.au}
N. Hurley-Walker,$^{1}$
G. Anderson,$^{1}$
K. Ross,$^{1,2}$
S. W. Duchesne$^{3}$
and T. J. Galvin$^{3}$
\\
$^{1}$International Centre for Radio Astronomy Research, Curtin University, Bentley WA 6102, Australia\\
$^{2}$Australian SKA Regional Centre (AusSRC), Curtin University, Kent Street, Bentley WA 6102, Australia\\
$^{3}$CSIRO Space and Astronomy, PO Box 1130, Bentley WA 6102, Australia
}

\date{Accepted XXX. Received YYY; in original form ZZZ} 
\pubyear{\the\year{}}

\begin{document}
\label{firstpage}
\pagerange{\pageref{firstpage}--\pageref{lastpage}}
\maketitle

\begin{abstract}
Image-based searches have become a complementary approach for identifying pulsars, particularly at MHz frequencies where scattering and high dispersion measures affect high-time resolution observations. In this work, we searched the Galactic Plane (GP) data release from the GaLactic and Extragalactic All-Sky Murchison Widefield Array eXtended (GLEAM-X) survey, a widefield continuum radio survey covering the $72 - 231$\,MHz frequency range, at the positions of known pulsars in the ATNF catalogue (version 2.6.2) that lie within its sky coverage. We present the spectral energy distribution for $193$ known pulsars located at $|b| < 11^{\circ}$. Notably, $106$ of these represent the first detections below $400$\,MHz. We also cross-match the GLEAM-X: GP compact source catalogue with unassociated sources in the Fermi Large Area Telescope (LAT) catalogue, filtering for gamma-ray spectral and variability properties, as well as coincidence with Active Galactic Nuclei (AGN). We have identified $106$ possible pulsar candidates. This work demonstrates the importance of sensitive, low-frequency Galactic plane surveys for detecting emission from known pulsars and presents a potential way of searching for new pulsar candidates that would otherwise be missed by traditional time-domain searches.
\end{abstract}


\section{Introduction}\label{sec:introduction}
Pulsars are rapidly rotating neutron stars that produce a stable sequence of radio pulses \citep[see ][ for a recent review]{Beskin2015}. Pulsars typically have a steep radio spectra with a mean spectral index $\alpha = -1.57$ \citep{Jankowski2018}, assuming a power law of the form $S_{\nu} \propto \nu^{\alpha}$ for the majority of the sources, with some fraction having spectral steepening (in the GHz regime) or low-frequency turnovers (below $200$\,MHz). The physics behind pulsar spectral behaviours is still poorly known. Low-frequency turnovers may be influenced by free-free absorption from ionised material within the pulsar structure or along the line of sight \citep{Kijak2009}, or by attenuation of the signal due to dispersion following $\nu^{-2}$ \citep{Kuniyoshi2015}.

Since their discovery in 1967, pulsars have been extensively studied using high-time resolution observations in order to detect and resolve the intricate structure of their pulses. This approach works well when employing single-dish telescopes at high radio frequencies ($\geq$~GHz). However, it becomes less effective at low radio frequencies, making a large portion of the pulsar population poorly explored below $400$\,MHz. One of the primary challenges at these frequencies is scattering, which increases steeply with decreasing frequency, following $\nu^{-4}$ \citep{Bhat2004,Geyer2017}. Additional factors that contribute to the difficulty in pulsar detections include dispersion, which causes a time delay in the signal leading to a broader pulse, the spectral turnover of pulsars at low frequencies, as well as the increase in system temperature due to Galactic synchrotron emission, which scales approximately as $\nu^{-2.6}$ \citep{Mozdzen2019}.

Recently, low-frequency studies of pulsars have gained attention. Rather than relying on traditional time-domain searches for periodic pulsations, the complementary approach of identifying steep-spectrum sources in images of wide-field continuum radio surveys has emerged. This method has proven effective with interferometric instruments such as the Giant Metrewave Radio Telescope \citep[GMRT][]{Swarup1991} and the Murchison Widefield Array \citep[MWA; ][]{Tingay2013,Wayth2018}. These have significantly expanded the number of known pulsars with flux density measurements at MHz frequencies \citep{Frail2016}. \cite{Sett2024} provide an example of this approach at $\leq 300$~MHz, detecting 83 known pulsars using both image-based and beamformed methods, $16$ of which were only detectable in Stokes~\textsc{i} images. Additionally, 14 pulsars were detected for the first time at low frequencies, previously missed by periodic searches due to having a high dispersion measure (DM) or being highly scattered.

Radio images have also been used in conjunction with gamma-ray catalogues to identify pulsars associated with previously unclassified sources. A notable example is given by \cite{Frail2018}, who utilised the Fermi 3FGL unassociated sources catalogue to guide their search for steep-spectrum radio sources potentially linked to pulsars. Based on gamma-ray characteristics, this targeted approach has proven to be significantly more efficient than traditional blind single-dish pulsation searches in uncovering new pulsar candidates, producing a list of $16$~candidates (seven of which later confirmed to have pulsation), including both millisecond pulsars (MSPs) and canonical pulsars.

Similar works, such as the GMRT-based search of steep-spectrum sources at $150$\,MHz within 3FGL confidence error ellipses, reinforced the utility of this approach, finding $11$~pulsar candidates characterised by bright radio emission in the MHz regime but faint at GHz frequencies, largely undetectable via conventional timing techniques \citep{Frail2016b}. Another notable example is the Transients and Pulsars with MeerKAT (TRAPUM) Large Survey Project, which, using gamma-ray spectral classification and targeted pointings in the range $856 - 1712$\,MHz, has discovered nine new MSPs out of $79$ Fermi-selected targets \citep{Clark2023}. These multi-wavelength approaches can significantly improve pulsar discovery rates.

Low-frequency detections are particularly valuable for probing pulsar emission mechanisms, investigating spectral turnover features caused by synchrotron self-absorption or thermal free-free absorption by gas present along the same line of sight of the source \citep[as described in][and references therein]{Swainston2021,Swainston2022}, and constraining statistical population properties such as DM and periodicity. Notably, \cite{Bilous2016} used LOw Frequency ARray (LOFAR) imaging observations in the $110-188$\,MHz range to estimate flux densities for $194$ pulsars. Furthermore, the inclusion of frequencies below $100$\,MHz was essential in characterising spectral behaviours, especially for identifying deviations from simple power-law spectra. A key advantage of continuum detections is the reliability of the measured flux densities. These measurements are averaged over long integration times relative to the pulses, and as such are less affected by variability intrinsic to the pulsar or propagation effects. This characteristic provides a more stable estimate of the source's average emission. However, averaging over the full pulse phase can obscure pulse components that originate from distinct regions of the pulsar magnetosphere and may exhibit varying widths or spectral behaviours, which inevitably introduces some loss in such measurements \citep{Vohl2024}. 
%Moreover, for pulsar observations at low frequencies, flux densities can be underestimated due to pulse broadening caused by scattering.

Although interstellar scintillation can strongly modulate pulsar flux densities at low frequencies, this effect is mitigated in the context of this work by the large temporal coverage of the continuum surveys, which effectively averages over scintillation-induced variability. For example, \cite{Murphy2017} detected $60$ known radio pulsars using the Galactic and Extragalactic All-sky MWA \citep[GLEAM; ][]{Wayth2015} survey images at $200$\,MHz. GLEAM observations, taken over two years, helped reduce scintillation effects on the sources. However, the identification of pulsars in GLEAM was highly limited by the sensitivity of the survey. On the other hand, the intensity fluctuations caused by interstellar scintillation can be advantageous in pulsar identification. This approach was demonstrated by \cite{Dai2016} using MWA observations. Their methodology involved the use of variance images to detect pulsars in radio continuum surveys, enabling searches across the entire field of view without being limited by the high computational cost associated with pixel-by-pixel techniques used in high time resolution searches. 

The GLEAM-eXtended \citep[GLEAM-X; ][]{Hurley2022,Ross2024} survey observed the same portion of the sky as GLEAM with twice the resolution and an order of magnitude improvement in sensitivity across the sky, importantly at Galactic latitudes. Here, we present a comparable analysis based on data from the Galactic Plane (GP) data release of the GLEAM-X survey (GLEAM-X: GP; Mantovanini et al., 2025, submitted), as well as a comparison with the unassociated gamma-ray sources that may reveal additional pulsar candidates worth following up in future studies.

The body of this paper is structured as follows: \sect~\ref{sec:jdgp} briefly summarises the GP data release of the GLEAM-X survey. \sect~\ref{sec:known} describes the methodology used to identify sources associated with the known pulsar population, including the spectral analysis performed. \sect~\ref{sec:high} outlines the approach employed to identify potential pulsar candidates through their possible association with gamma-ray sources.\sects~\ref{sec:results} and \ref{sec:conclusion} provide a description of the key results from both approaches and conclude with a brief overview.

\section{GLEAM-X: GP data release} \label{sec:jdgp}
The MWA has conducted extensive mapping of the sky South of Declination $+30^{\circ}$, resulting in two major continuum surveys. The instrument Phase~\textsc{i} configuration observed the sky over 28 nights between 2013 and 2014, with additional observations in the following year to replace data affected by poor ionospheric conditions. These observations formed the basis of the GLEAM survey, which has a spatial scale sensitivity of $ 2' - 15^{\circ}$, reaching typical noise levels of $10$~\mJypbm in the extragalactic sky \citep{Hurley2017} and 50 -- 100~\mJypbm along the GP \citep{Hurley2019c}. 

In contrast, the ``extended'' Phase~\textsc{ii} configuration of MWA was used for GLEAM-X observations, taking data over $113$ nights from 2018 to 2020. These observations are incrementally made available to the community as regions are completed. This configuration enabled higher resolution imaging, capturing smaller spatial scales ($45^{''} - 20'$) and achieving improved sensitivity over long integrations ($\sim 1$~\mJypbm) thanks to the lowered confusion limit. 

The data used in this analysis are taken from the GLEAM-X: GP data release. GLEAM-X: GP combines observations from GLEAM and GLEAM-X surveys through joint deconvolution, achieving higher sensitivity to a broad range of spatial scales. The survey spans a $72 - 231$\,MHz frequency range, divided into $20$ sub-bands of $7.68$\,MHz. The data release covers $\approx 3800$~deg$^2$ of the southern GP, specifically from longitudes $233^{\circ} < l < 44^{\circ}$ and latitudes $|b| < 11^{\circ}$. This wide spatial and spectral coverage enables accurate flux density measurements even for extended sources. Mosaics are made for each sub-band, along with an additional wideband image combining the top eight frequency sub-bands ($170 - 231$\,MHz) for a more sensitive source-finding approach. The resulting catalogue contains 98,207 elements, each measured across the $20 \times 7.68$\,MHz frequency sub-bands and with source position accuracy within $3$ arcseconds.

We have selected a subset of this catalogue to consider only compact objects. The compactness criteria have been defined following equation $2$ in \cite{Meyers2017} $(\chi - 3 \Delta \chi) \geq 1$, for which: 

\begin{equation}
    \chi = \dfrac{ab}{a_\mathrm{PSF}b_\mathrm{PSF}} 
\end{equation}
%&\Delta \chi \simeq \sqrt{\left( \dfrac{\Delta a}{a} \right)^2 + \left( \dfrac{\Delta b}{b} \right)^2}

where $a_\mathrm{PSF}$ and $b_\mathrm{PSF}$ are the major and minor axes for the local point spread function (PSF), while $a$ and $b$ correspond to the extent of the source. The error on the compactness ($\Delta \chi$) is calculated by summing the fractional errors in $a$ and $b$ in quadrature.

\section{Known pulsars} \label{sec:known}
We begin by focusing on the known population of pulsars, aiming to identify counterparts in the GLEAM-X: GP catalogue in order to provide accurate flux density measurements at MHz frequencies, an observational regime that has been relatively unexplored. In the following sections, we describe the data selection and process and detail the analysis conducted to investigate the spectral properties of the detected sources.

\subsection{Sample selection}
We used the Australia Telescope National Facility (ATNF) pulsar catalogue v2.6.2 \citep[][]{Manchester2005} \footnote{https://www.atnf.csiro.au/research/pulsar/psrcat/} to perform a cross-match with the GLEAM-X: GP catalogue to identify potential associations. We constructed an error ellipse centred on the RA and Dec using the reported positional uncertainties in those coordinates for each pulsar in the ATNF pulsar catalogue. Similarly, each source from the GLEAM-X: GP catalogue was assigned an elliptical region defined as the beam size aligned along the beam orientation. We considered a match to occur when the error ellipse of a pulsar and a source overlapped. 

\begin{figure}
\centering
\includegraphics[width=1.0\linewidth]{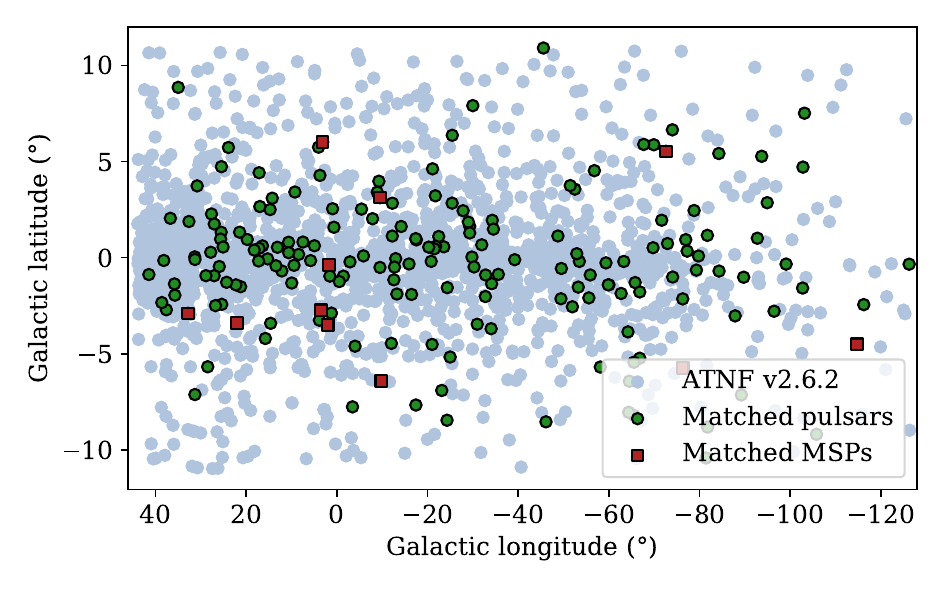}
\caption{Distribution of known pulsars from the ATNF Pulsar Catalogue (light blue dots) and those detected in GLEAM-X: GP (green circles), as presented in this work. MSPs are highlighted in red squares.} \label{fig:matched}
\end{figure}

The cross-match resulted in $608$~associations, with $13$ pulsars linked to multiple GLEAM-X: GP counterparts. All the matches were then visually inspected in the wide-band ($170 - 231$\,MHz) image of GLEAM-X: GP, ruling out artefacts, and part of diffuse structures that have erroneously been classified as positive matches. We also inspected the lowest wavelength band of the Widefield Infrared Survey Explorer \citep[AllWISE;] []{Wright2010,Mainzer2011} images at $3.4$\;$\mu$m to rule out the coincidence with a globular cluster or Active Galactic Nuclei (AGN). This selection resulted in $426$ associations corresponding to $202$ distinct pulsars. Of these, two pulsars have 2--4 matches each, while seven have more than $5$ matches due to large errors in the pulsar position, in some cases resulting in up to $80$~matches for a single pulsar. We excluded these seven pulsars from further analysis, as it is not possible to confidently identify the true counterpart. We leave in the analysis the two pulsars with fewer matches, which can be more easily explored. The remaining matched pulsars are shown in \fig~\ref{fig:matched}.

\begin{figure}
\centering
\includegraphics[width=1.0\linewidth]{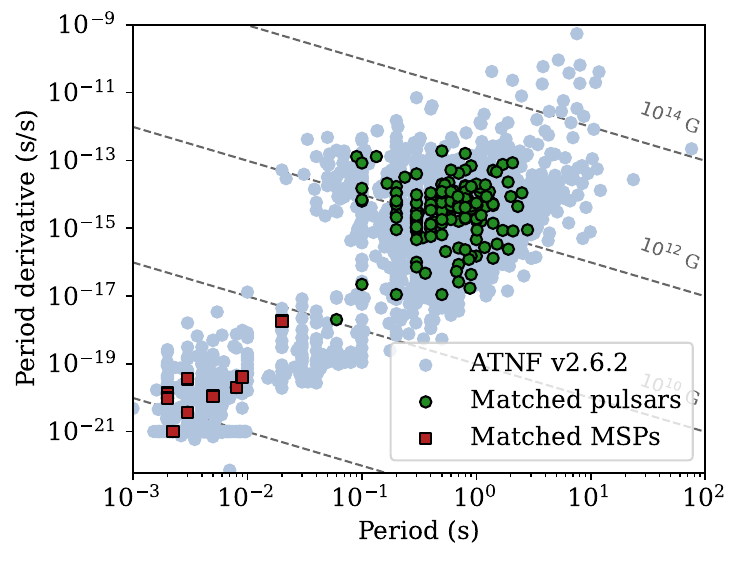}
\caption{$P-\dot{P}$ diagram showing the distribution of known pulsars from the ATNF Pulsar Catalogue (light blue dots) and those detected in GLEAM-X: GP (green circles). MSPs are highlighted in red squares. All the elements with no period derivative value have been set to $10^{-21}$. The surface magnetic field intensity contours are approximated as $3.2 \times 10^{19} G \sqrt{P \dot{P}}$ and reported as black dashed lines.} \label{fig:ppdot}
\end{figure}

$183$ of the cross-matched sources were classed as canonical pulsars in the ATNF catalogue. Additionally, $10$ matches were classed as MSPs, characterised by their short spin periods (less than $30$ milliseconds) and extremely low period derivatives (spin-down rates), specifically below $10^{-16}$\,s\,s$^{-1}$. One additional source (J1751$-$2737) lacks a measured period derivative but exhibits a very short spin period of $2$ milliseconds. \fig~\ref{fig:ppdot} shows the relevant $P-\dot{P}$ diagram.

To account for the possibility of chance alignments, we evaluated the probability of spurious associations based on the source density of the GLEAM-X: GP catalogue. With $\approx 98,000$ sources spread across $\approx 3,800$ square degrees, the average surface density is approximately 26 sources per square degree. Using an average search ellipse of $0.9 \times 1.8$-arcminutes around each target, this density yields a random match probability of about $4\%$. Applied to our sample of $193$ pulsars, this suggests that roughly $7$ may be coincidental rather than physically meaningful. This estimate may be conservative as the analysis focused on areas near the Galactic plane, where the background source density is typically elevated. Despite this factor potentially increasing the chance of spurious associations, we expect the match probability to remain relatively low (on the order of $4\%$), and thus the number of false matches to be limited to only a small fraction of the total sample.

\begin{figure}
\centering
\includegraphics[width=1.0\linewidth]{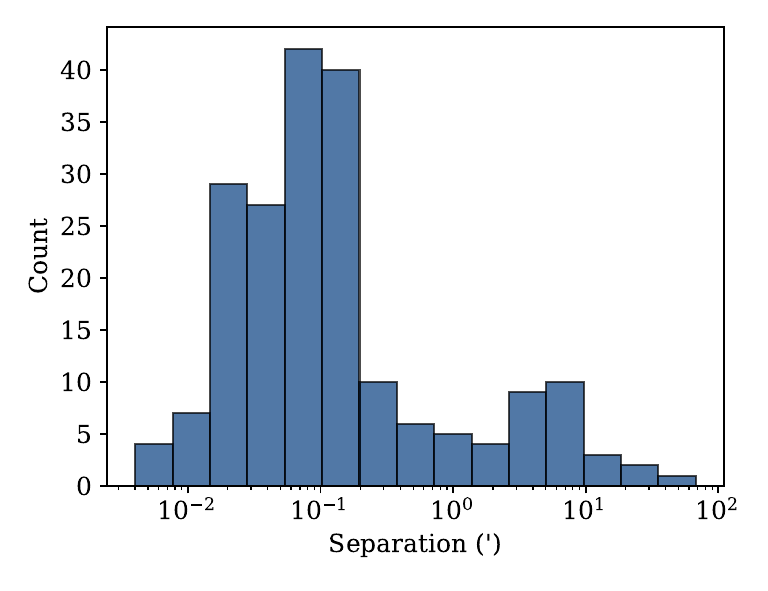}
\caption{Histogram of angular separations between matched ATNF pulsars and GLEAM-X: GP sources, reported in arcminutes. The x-axis is shown on a logarithmic scale.} \label{fig:sep}
\end{figure}

For each match, the angular separation between the ATNF pulsar and the corresponding GLEAM-X: GP source was calculated, and the resulting distribution is shown in \fig~\ref{fig:sep}. The majority of matches exhibit small separations, indicating good positional agreement. The tail at larger separations is primarily due to the significant positional uncertainties associated with pulsars in the ANTF catalogue. The overall close alignment proves good agreement between the two datasets and is crucial for confirming the identity of sources and reducing the likelihood of false positives, especially in regions with high source density.

\subsection{Spectral fitting} \label{sec:software}
It is important to determine the flux density measurements of pulsars to better understand their spectral properties and underlying emission mechanisms, which remain poorly constrained. Free-free absorption from ionised material and/or attenuation of the signal due to dispersion may play a key role in shaping their spectra. \cite{Jankowski2018} studied the spectral properties of $441$ pulsars in order to model pulsar spectra and gain insights into the dominant emission processes by observing the sources using the Murriyang, CSIRO's Parkes radio telescope (hereafter Parkes) at $728$\,MHz, $1382$\,MHz, and $3100$\,MHz and combining data from the literature. The spectra obtained were more diverse than previously thought and could not be adequately described by a simple power-law model, indicating the need for additional components.

Based on this work, \cite{Swainston2022} designed the \textit{pulsar$\_$spectra} software package specifically for modelling pulsar flux density measurements across a broad frequency range. The software supports multiple spectral models - including single power-law, broken power-law, and low frequency turnovers - allowing flexible treatment of diverse spectral behaviours. For each pulsar, \textit{pulsar$\_$spectra} takes as input a set of flux density measurements at various frequencies, along with their associated uncertainties. It then applies a Bayesian framework to estimate the best-fitting model parameters, employing Markov Chain Monte Carlo (MCMC) sampling via the emcee backend to explore the parameter space. The output includes best-fit parameter values and diagnostic plots to assess the quality of the fits. 

While pulsar spectra have been relatively well studied in the $400-1400$~MHz range, their behaviour at lower frequencies remains less explored, primarily due to challenges such as flux variability from interstellar scintillation, particularly for pulsars with low DMs. Our work addresses this gap by providing flux density measurements from GLEAM-X: GP in $20$~subbands for each matched source and fitting these using the default settings of the \textit{pulsar$\_$spectra} software.

For each pulsar, the software compiles previously published flux density measurements from the literature (saved in an open-source catalogue) and fits spectral models using these data. We included in the calculation the GLEAM-X: GP flux measurements and their associated uncertainties for each radio source matched to a given pulsar. Five spectral models were fitted to the combined data: a simple power-law, a broken power-law, a double turnover power-law, a low frequency turnover power-law, and a high frequency cutoff power-law. All the models were used in previous literature studies and are physically motivated. For example, a low-frequency turnover may be observed in cases of free-free absorption along the line of sight or a decrease in the source emission at low frequencies. The best fit model is selected by performing the Akaike information criterion (AIC), which balances how well the model has been fitted to the data and the model complexity to avoid overfitting.

It should be noted that \textit{pulsar$\_$spectra} relies on an older version of the ATNF Pulsar Catalogue. Consequently, eight of the newly matched sources in our sample are not recognised by the software. No flux density measurements are available for these pulsars in the latest catalogue (v2.6.2), because only lower limits are listed in the discovery paper \citep{Sengar2025}. For this reason, spectral energy distributions (SEDs) are constructed with GLEAM-X: GP data only.
%In these cases, we saved flux density values directly from the ATNF catalogue v2.6.2, if available, along with their errors. If the error on the flux is not given, we arbitrarily set it to $1$\,mJy.

\begin{figure*}
    \centering
    \subfloat[][\emph{Simple power-law}]
    {\includegraphics[scale=0.7]{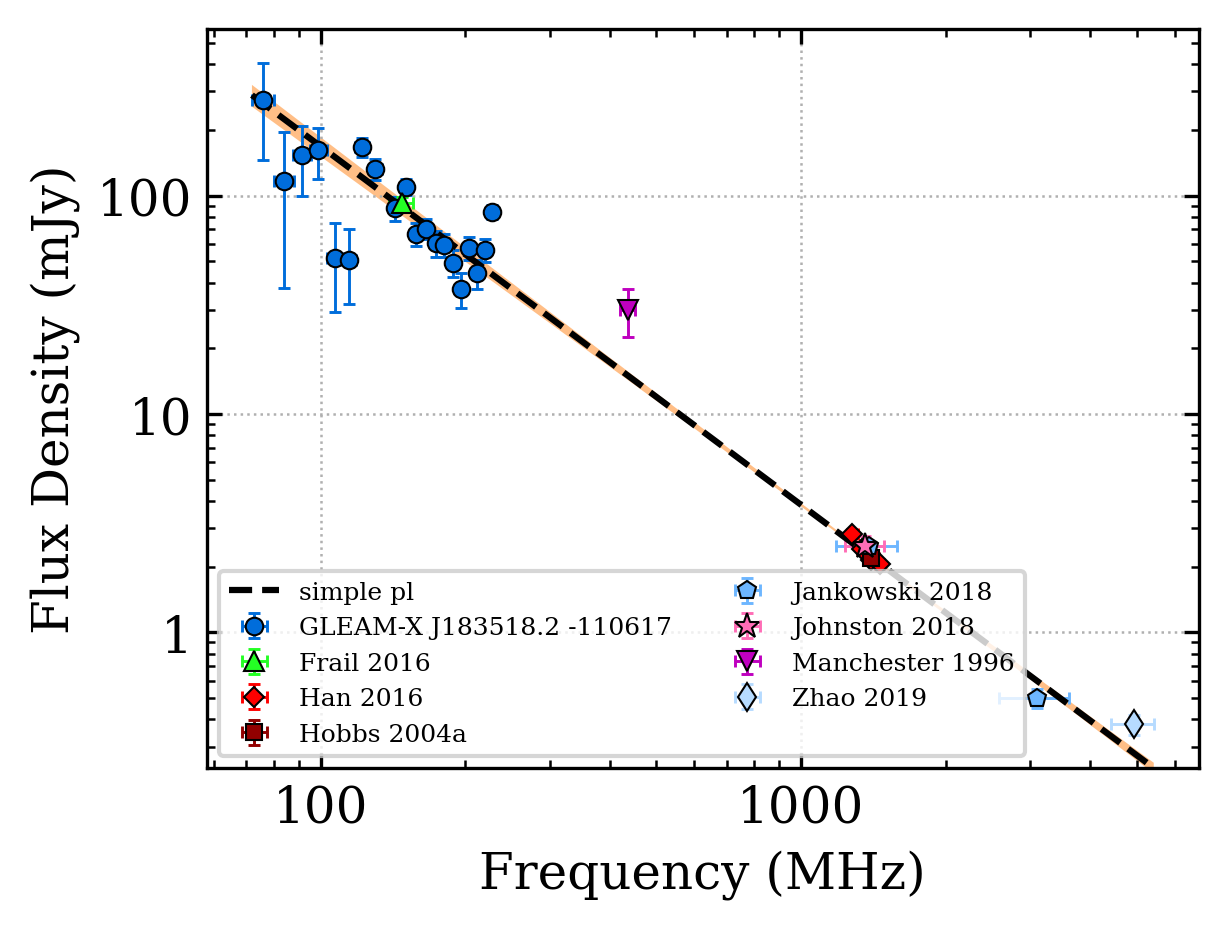}} \quad 
    \subfloat[][\emph{Broken power-law}]
    {\includegraphics[scale=0.7]{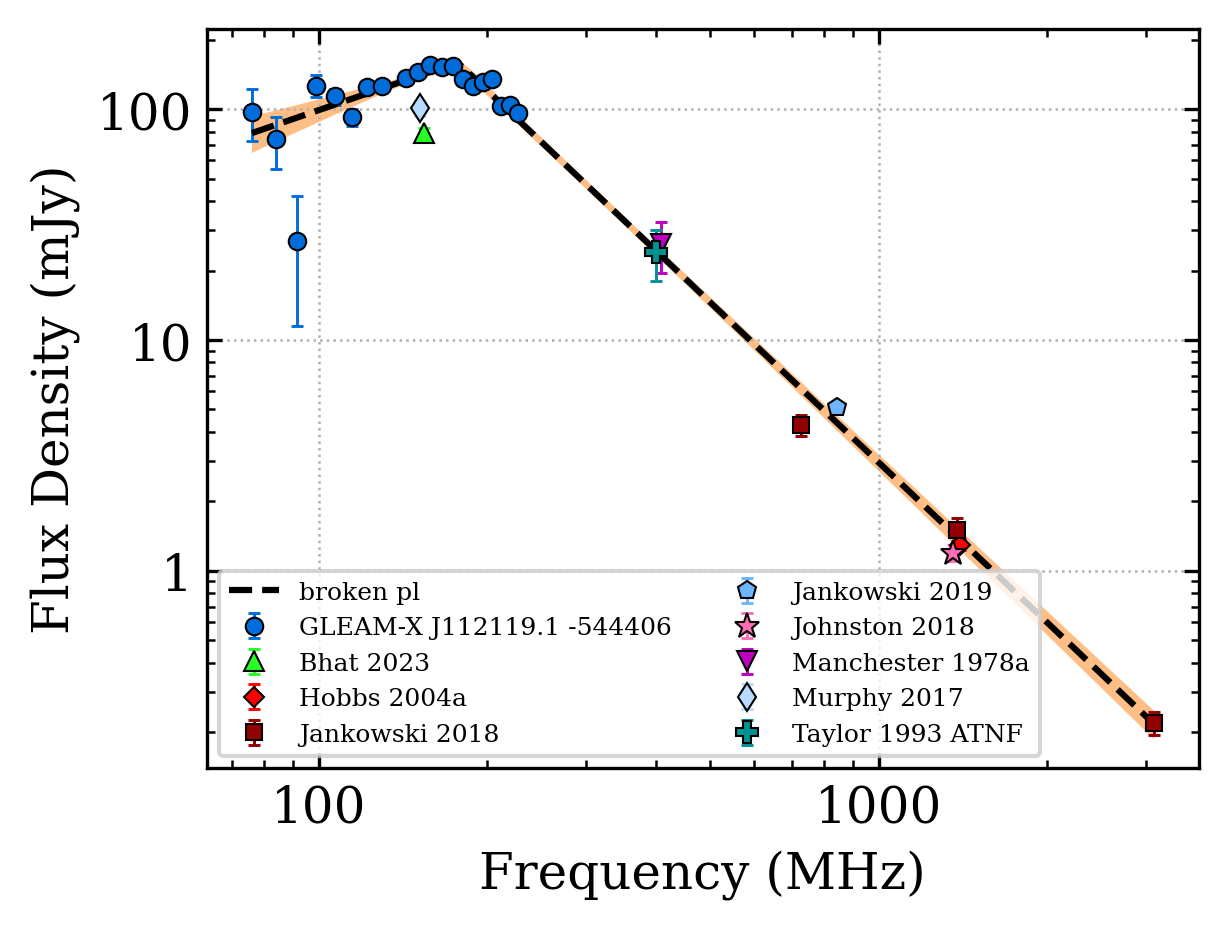}} \\
    
    \subfloat[][\emph{Low frequency turnover power-law}]
    {\includegraphics[scale=0.7]{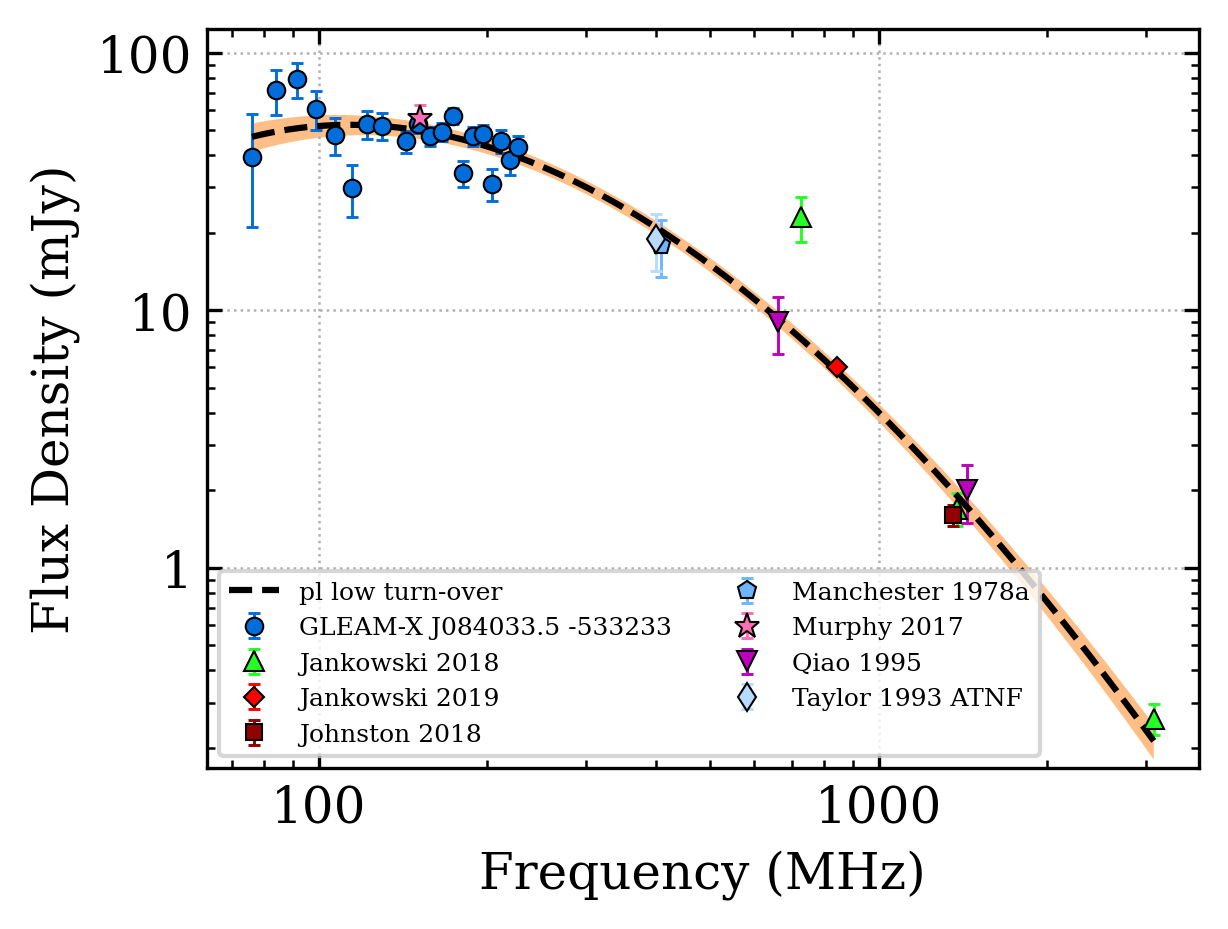}} \quad
    \subfloat[][\emph{High frequency cutoff power-law}]
    {\includegraphics[scale=0.7]{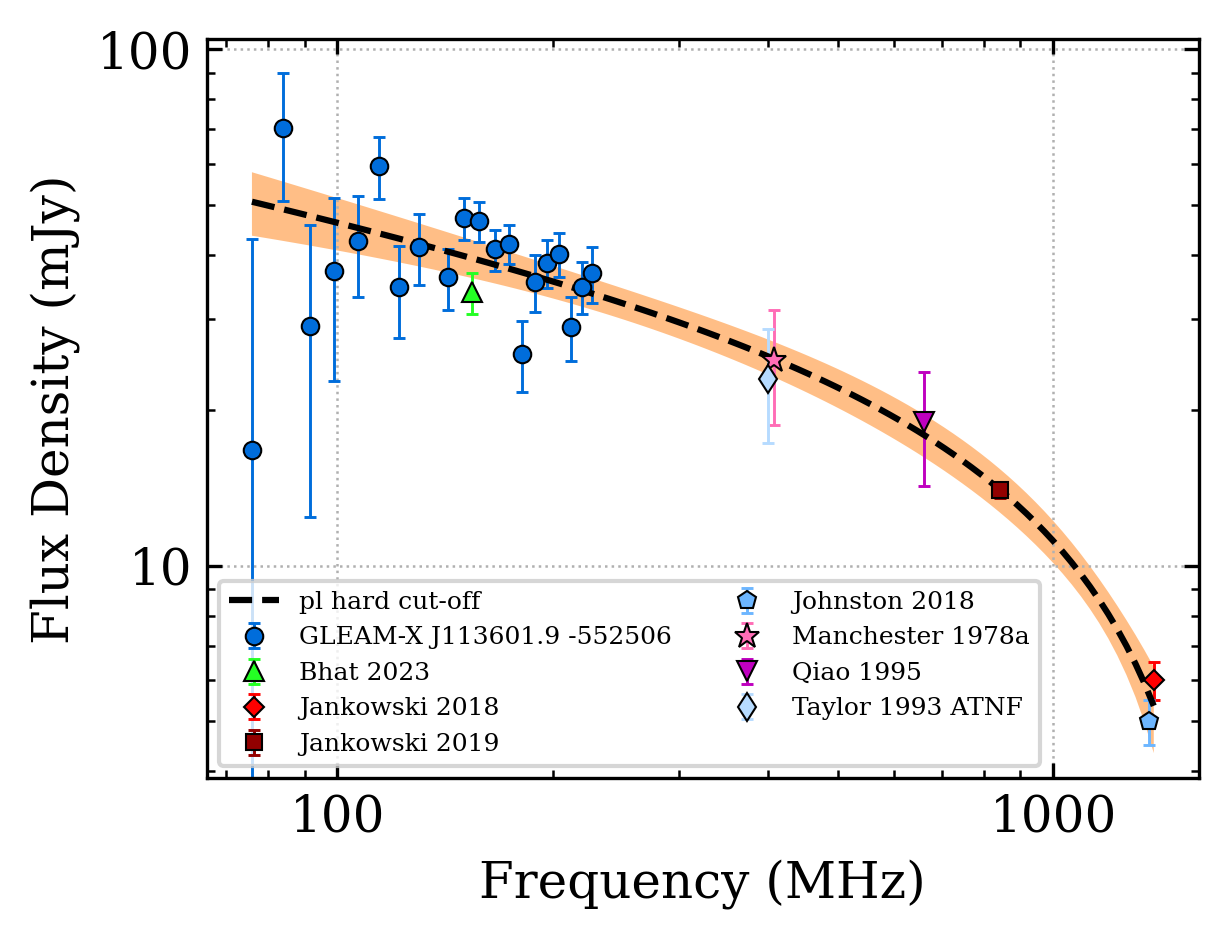}} \\
    
    \subfloat[][\emph{Double turnover power-law}]
    {\includegraphics[scale=0.7]{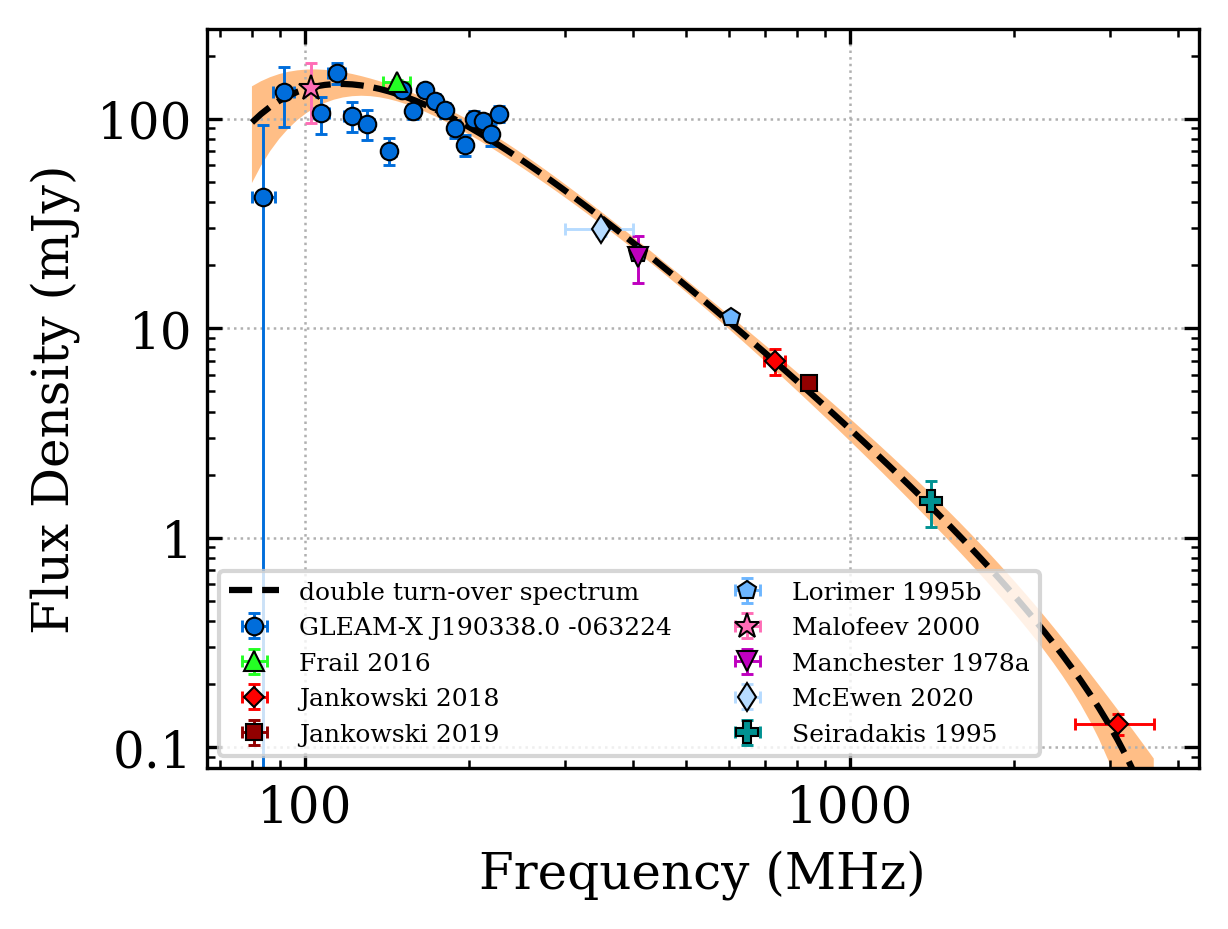}} \\
    
    \caption{Example SED fit for pulsars using the five models available in the \textit{pulsar$\_$spectra} software package: power-law (a), broken power-law (b), low-frequency turnover power-law (c), high frequency cutoff power-law (d), and double turnover power-law (e). Blue circles represent data points for flux densities measured in this work with associated uncertainties.} \label{fig:fit-models}
\end{figure*}

An example fit with all five models employed is reported in \fig~\ref{fig:fit-models}.

\section{High energy candidates} \label{sec:high}
%\subsection{Radio-quiet pulsars}
% Check radio quiet pulsars in the Fermi catalogue
Image-based approaches have also been increasingly used in the identification of potential gamma-ray pulsars, serving as a complement to more traditional methods based on pulsation searches. In the following, we describe our methodology for identifying promising radio candidates that may be associated with gamma-ray unidentified sources that exhibit pulsar-like properties at high energies.

\subsection{Fermi unassociated sources}
We used the fourth catalogue \citep[4FGL-DR4,][]{Abdollahi2022,Ballet2023} of the Large Area Telescope \citep[LAT][]{Atwood2009} sources, based on $14$ years of survey data (August 2008 -- August 2022) over the energy range $50 - 1000$\,MeV and containing $7,194$ sources, to perform a cross-match with the GLEAM-X: GP catalogue in order to find potential pulsar candidates worth following up in future pulsation searches. 

This is achieved by combining a positional cross-match with a prior classification of unassociated 4FGL-DR4 sources based on their spectral and variability gamma-ray properties, followed by a visual inspection of the resulting matches. We considered only the 4FGL catalogue containing compact sources~\footnote{https://fermi.gsfc.nasa.gov/ssc/data/access/lat/14yr$\_$catalog/}. Pulsars tend to have relatively steep spectra with pronounced curvature and small variability \citep{Zhu2024}. We therefore filtered this catalogue to pick out pulsar-like sources, requiring each source to satisfy all of the following criteria: 
%(gll$\_$psc$\_$v35.fit)
\begin{itemize}
    \item Moderate curvature (logarithm of the log parabolic fit significance $0-2.0$): Pulsars typically show a curved spectrum (modelled either by a log-parabola or an exponentially cutoff power-law) across the Fermi-LAT energy range. The curvature significance is quantified by comparing how well a curved model fits the data compared to a simple power-law, and it represents the number of standard deviations by which the curved model improves the fit. A high significance implies strong evidence for curvature \citep{Abdollahi2020}. The logarithm of this quantity is often used in classification with a value of $1$ to indicate strong curvature ($10 \sigma$ improvement), and values below $0.5$ ($\approx 3 \sigma$) are considered insignificant, suggesting a power-law fit is sufficient. Pulsars have an average value of $1$, reflecting that their curvature is statistically significant \citep{Saz2016}. AGN, on the other hand, have low curvature significance (below $0.4$) and are well-described by a simple power-law \citep{Ajello2020}.
    \item Photon index $1.8-3.0$: The photon index characterises the slope of a source's energy spectrum when approximated by a simple power-law with high values ($>2.3$), indicating a softer (steeper) spectrum. Although curved models better describe pulsar spectra, this parameter is quantitatively useful as a proxy for spectral steepness, and it is used in source classification, particularly to distinguish pulsars from AGN. Most AGN are characterised by a photon index in the range of $1.5-2.2$ \citep{Singal2015}. However, there is some overlap; the parameter helps reduce the classification ambiguity when combined with, e.g. the curvature significance.
    \item Low variability (log variability index $0.4-1.7$): Pulsars present a regular and stable gamma-ray emission pattern over months to years unless subject to changes in the emission geometry. This behaviour results in a low variability index, calculated by comparing the average flux over the full catalogue time interval to the flux in $1$-year bins \citep{Abdollahi2020}. The logarithm of this quantity is used in source classification, with values $< 1$ corresponding to non-variable objects. The distribution of the parameter for pulsars has a peak around $0.9-1.2$. In contrast, AGN have strong gamma-ray variability with indices greater than $2$ \citep{Ajello2020}, mainly due to particle acceleration in their jets.
\end{itemize}

After applying these cuts, our working sample consisted of unassociated 4FGL compact sources that have pulsar-like photon indices, low variability, and evidence for spectral curvature. We then cross-matched the filtered gamma-ray sources against the GLEAM-X: GP compact catalogue. For each candidate, we considered the $95\%$ error ellipses of the 4FGL filtered sources and assigned an elliptical region defined as the beam size aligned along the beam orientation for each GLEAM-X: GP compact source. We considered a match to occur when the position of a GLEAM-X: GP source falls within the ellipse of a gamma-ray source. 

The cross-match resulted in $591$ preliminary associations, with most of the gamma-ray entries linked to multiple radio ellipses. Many of the matches occur in complex regions or may be chance coincidences. We therefore applied further filtering to clean the list. First, we removed any gamma-ray source with a high number of matches (more than $5$), typically caused by a large uncertainty in source position or by its placement in very crowded fields and hence likely unreliable. Second, we filtered the associations by looking at the flags assigned to each source in the 4FGL-DR4 catalogue. We excluded flags denoted by the bits ``1'', ``2'', ``5'', and ``6''~\footnote{The flags are encoded in the 4FGL-DR4 catalogue as individual bits within a single integer column. Each flag n corresponds to a specific bit in the binary representation of that value, and its value is given by: $2^{(n-1)}$.}. Any source flagged in this way was considered improbable because it implied localisation errors or background/modelling uncertainties. 

\begin{figure}
\centering
\includegraphics[width=1.0\linewidth]{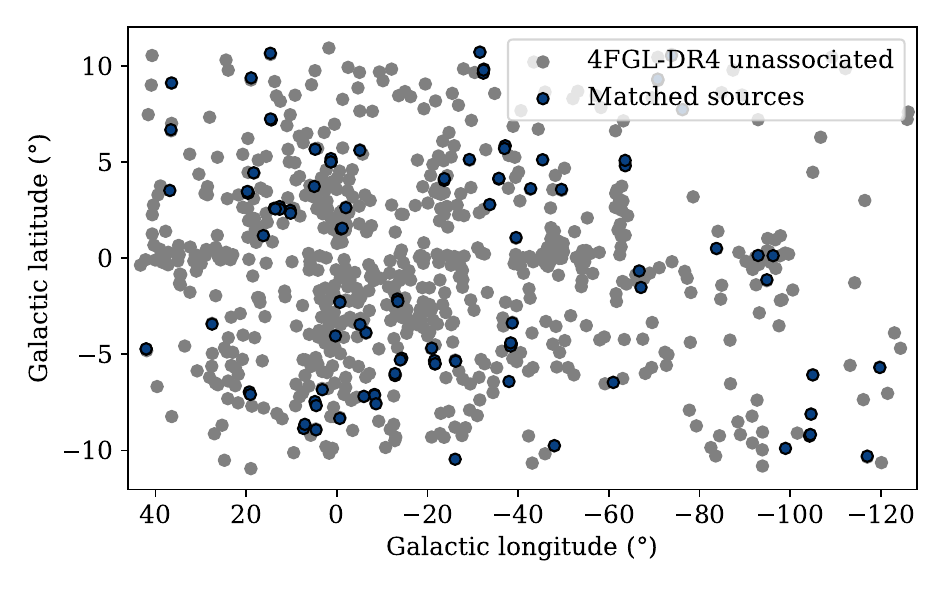}
\caption{Distribution of the unassociated compact sources from the 4FGL-DR4 Catalogue (light grey dots) and potential associated radio sources detected in GLEAM-X: GP (blue circles).} \label{fig:gamma}
\end{figure}

This reduced the associations to $191$~corresponding to $117$~unique gamma-ray sources. All the remaining matches were visually inspected in the wide-band ($170 - 231$\,MHz) image of the GLEAM-X: GP data release, ruling out artefacts, and part of diffuse structures that were erroneously classified as positive matches. We also queried the online SIMBAD \citep{Wenger2000} portal to rule out the coincidence with AGN that have not been removed by the previous steps. This selection resulted in $73$ unique gamma-ray sources with a total of $106$ radio associations. The distribution of the resulting sample is shown in \fig~\ref{fig:gamma}

We inspected the Parkes telescope data portal~\footnote{https://data.csiro.au/domain/atnf} to verify whether the positions of the matched gamma-ray sources had been included in previous pulsar search campaigns. Our investigation revealed that $33$ gamma-ray sources had been previously observed, with pointings mostly directed at the coordinates of the gamma-ray ellipses. \Cref{tab:unassoc,tab:observed} list the source names and their coordinates (in degrees) for those matched in both the 4FGL-DR4 and GLEAM-X: GP catalogues, distinguishing between sources not previously observed and those that have been targeted in earlier projects. 

The majority of the Parkes projects observed the corresponding gamma-ray sources prior to 2022. As such, if a pulsar had been detected, it would likely have been reported in a publication and subsequently included in either the ATNF pulsar catalogue or the 4FGL-DR4 catalogue, depending on the nature of the discovery. Only four of the matched sources were part of more recent observing programs. Project P1054 targeted source J1208.0$-$6900 as part of TRAPUM, but the discovery is not listed on the TRAPUM discovery website~\footnote{https://www.trapum.org/discoveries/}. Project P1194 aimed to identify MSPs among Fermi-selected candidates, but no relevant publication or result could be located. Project P1211 focused on MSPs towards steep-spectrum radio sources, though no clear association with gamma-ray source J1517.9$-$5233 was found. Lastly, project P1348 observed 1815.8$-$1416 with the goal of confirming its nature as an eclipsing binary. While the associated paper is in preparation (Petrou et al.), preliminary results show evidence of radio-band periodicity, although no pulsations have been detected to date. This source remains a promising candidate, and further follow-up observations need to be undertaken.

The position of the telescope pointing can affect the final result of a pulsation search. Knowing the exact location of the expected pulsar enables targeted observations, which are far more sensitive than blind searches over larger areas. The radio sources matched to Fermi error ellipses have the potential to provide more precise localisation, enabling the telescope beam to be accurately centred towards the most likely coordinates in the sky, and so improving the sensitivity to pulsed emission by maximising the signal-to-noise ratio.

\begin{table*}
\centering
\caption{Source names and coordinates of all the gamma-ray sources associated with at least one GLEAM-X: GP radio source, excluding those previously targeted by Parkes pulsar search campaigns. For each gamma-ray source, the table lists the name (column $1$), the sky coordinates (RA and Dec, columns $2$ and $3$) of the ellipse centre, spectral type (either power-law [PL] or log parabola [LP]), photon index $\Gamma$ (defined as $N(E) \propto (E/E_0)^{-\Gamma}$), and energy flux at $100$\,MeV. For each associated radio source (name in column $7$), the table provides the coordinates (columns $8$ and $9$) and flux density at $200$\,MHz.} \label{tab:unassoc}
\begin{tabular}{lccccccccr}
    \hline
    Name & RA & Dec & Spec & Photon & E$_{100} \times 10^{-12}$ & Name & RA & Dec & \multicolumn{1}{c}{S$_\mathrm{200 MHz}$}\\
    4FGL- & $^{\circ}$ & $^{\circ}$ & Type & Index & erg cm$^{-2}$ s$^{-1}$ & GLEAM-X- & $^{\circ}$ & $^{\circ}$ & \multicolumn{1}{c}{mJy} \\
	\hline
    J0708.8$-$3121 & 107.211 & $-$31.365 & LP & 2.79 & 2.14 & J070904.7$-$312121 & 107.27 & $-$31.356 & 23 \\
    J0741.9$-$4157 & 115.491 & $-$41.963 & PL & 2.67 & 4.11 & J074202.6$-$420429 & 115.511 & $-$42.075 & 53 \\
    J0741.9$-$4157 & 115.491 & $-$41.963 & PL & 2.67 & 4.11 & J074207.9$-$415634 & 115.533 & $-$41.943 & 557 \\
    J0741.9$-$4157 & 115.491 & $-$41.963 & PL & 2.67 & 4.11 & J074155.7$-$415221 & 115.482 & $-$41.873 & 290 \\
    J0753.8$-$4700 & 118.451 & $-$47.0061 & PL & 2.60 & 2.79 & J075306.2$-$465803 & 118.276 & $-$46.968 & 278 \\
    J0848.8$-$43.28 & 132.211 & $-$43.4731 & LP & 2.91 & 21.4 & J084858.9$-$433300 & 132.245 & $-$43.55 & 141 \\
    J0900.2$-$4608 & 135.051 & $-$46.1347 & PL & 2.50 & 4.19 & J090104.7$-$460408 & 135.27 & $-$46.069 & 63 \\
    J0942.1$-$5215 & 145.5484 & $-$52.2652 & LP & 2.33 & 2.43 & J094238.5$-$521818 & 145.660 & $-$52.305 & 37 \\
    J1048.4$-$5030 & 162.107 & $-$50.5132 & PL & 1.85 & 2.32 & J104824.3$-$502939 & 162.101 & $-$50.494 & 417 \\
    J1109.1$-$4853 & 167.278 & $-$48.8967 & LP & 2.29 & 1.87 & J110912.7$-$485816 & 167.303 & $-$48.971 & 36 \\
    J1122.4$-$6237 & 170.6085 & $-$62.6325 & PL & 2.31 & 8.72 & J112233.9$-$623901 & 170.641 & $-$62.650 & 37 \\
    J1123.2$-$5111 & 170.8225 & $-$51.1956 & LP & 2.29 & 8.72 & J112325.8$-$511502 & 170.857 & $-$51.251 & 21 \\
    J1123.2$-$5111 & 170.8225 & $-$51.1956 & LP & 2.29 & 8.72 & J112345.6$-$511149 & 170.940 & $-$51.197 & 39 \\
    J1127.9$-$6158 & 171.9945 & $-$61.9807 & LP & 2.60 & 9.57 & J112806.9$-$615750 & 172.029 & $-$61.964 & 53 \\
    J1349.1$-$5829 & 207.283 & $-$58.4893 & PL & 2.68 & 7.58 & J134915.1$-$582626 & 207.313 & $-$58.440 & 16 \\ 
    J1415.2$-$5550 & 213.8166 & $-$55.8389 & LP & 2.85 & 3.89 & J141536.9$-$554930 & 213.903 & $-$55.825 & 41 \\
    J1437.6$-$5616 & 219.4096 & $-$56.2782 & LP & 2.62 & 2.70 & J143657.7$-$561723 & 219.240 & $-$56.290 & 21 \\
    J1443.7$-$7037 & 220.946 & $-$70.621 & PL & 2.44 & 5.04 & J144315.3$-$703747 & 220.814 & $-$70.630 & 63 \\
    J1529.4$-$6027 & 232.367 & $-$60.4583 & LP & 2.62 & 4.87 & J153007.4$-$602653 & 232.531 & $-$60.448 & 26 \\
    J1537.3$-$6110 & 234.334 & $-$61.1771 & LP & 2.43 & 2.75 & J153635.5$-$611238 & 234.148 & $-$61.211 & 49 \\
    J1537.3$-$6110 & 234.334 & $-$61.1771 & LP & 2.43 & 2.75 & J153724.8$-$611441 & 234.353 & $-$61.245 & 36 \\
    J1537.3$-$6110 & 234.334 & $-$61.1771 & LP & 2.43 & 2.75 & J153817.0$-$611252 & 234.571 & $-$61.214 & 58 \\
    J1537.3$-$6110 & 234.334 & $-$61.1771 & LP & 2.43 & 2.75 & J153715.5$-$610556 & 234.315 & $-$61.099 & 136 \\
    J1547.4$-$4802 & 236.8501 & $-$48.0393 & LP & 2.56 & 9.34 & J154730.8$-$475903 & 236.878 & $-$47.984 & 45 \\
    J1550.3$-$6223 & 237.59 & $-$62.391 & PL & 2.67 & 3.38 & J155013.8$-$622716 & 237.558 & $-$62.454 & 59 \\
    J1616.0$-$4501 & 244.009 & $-$45.0213 & PL & 2.50 & 5.49 & J161552.7$-$450929 & 243.97 & $-$45.158 & 203 \\
    J1616.0$-$4501 & 244.009 & $-$45.0213 & PL & 2.50 & 5.49 & J161525.8$-$450623 & 243.857 & $-$45.106 & 73 \\
    J1620.5$-$5729 & 245.128 & $-$57.4945 & LP & 2.47 & 3.81 & J162026.4$-$572935 & 245.11 & $-$57.493 & 134 \\
    J1647.5$-$5319 & 251.899 & $-$53.332 & LP & 2.77 & 3.52 & J164734.4$-$532357 & 251.893 & $-$53.399 & 182 \\
    J1647.5$-$5319 & 251.899 & $-$53.332 & LP & 2.77 & 3.52 & J164732.3$-$532629 & 251.885 & $-$53.441 & 69 \\
    J1705.4$-$4850 & 256.366 & $-$48.8339 & LP & 2.74 & 5.49 & J170458.4$-$485424 & 256.243 & $-$48.907 & 49 \\
    J1730.1$-$4343 & 262.539 & $-$43.727 & LP & 2.83 & 5.01 & J172923.5$-$434850 & 262.348 & $-$43.813 & 45 \\
    J1730.1$-$4343 & 262.539 & $-$43.727 & LP & 2.83 & 5.01 & J173047.9$-$433537 & 262.7 & $-$43.593 & 108 \\
    J1730.3$-$2913 & 262.577 & $-$29.2302 & LP & 2.28 & 2.19 & J173032.8$-$291250 & 262.637 & $-$29.214 & 93 \\
    J1735.2$-$2153 & 263.8187 & $-$21.8859 & PL & 2.50 & 4.78 & J173527.2$-$215300 & 263.863 & $-$21.883 & 88 \\
    J1737.1$-$2901 & 264.276 & $-$29.0272 & PL & 2.54 & 11.6 & J173727.4$-$285525 & 264.364 & $-$28.923 & 87 \\
    J1737.1$-$2901 & 264.276 & $-$29.0272 & PL & 2.54 & 11.6 & J173640.5$-$290700 & 264.169 & $-$29.116 & 137 \\
    J1739.1$-$1059 & 264.7836 & $-$10.9948 & LP & 2.61 & 3.99 & J173857.8$-$105640 & 264.741 & $-$10.944 & 225 \\
    J1742.8$-$2246 & 265.724 & $-$22.7718 & LP & 2.54 & 7.11 & J174259.6$-$224430 & 265.749 & $-$22.742 & 45 \\
    J1750.8$-$1246 & 267.7245 & $-$12.7802 & LP & 2.57 & 3.85 & J175047.7$-$125215 & 267.699 & $-$12.871 & 76 \\ 
    J1750.8$-$1246 & 267.7245 & $-$12.7802 & LP & 2.57 & 3.85 & J175057.3$-$124154 & 267.739 & $-$12.698 & 83 \\
    J1752.4$-$0758 & 268.1007 & $-$7.9776 & LP & 2.84 & 3.88 & J175155.9$-$075735 & 267.983 & $-$7.960 & 321 \\
    J1755.9$-$4009 & 268.976 & $-$40.1513 & LP & 2.18 & 1.31 & J175606.3$-$401225 & 269.026 & $-$40.207 & 55 \\
    J1805.9$-$1549 & 271.492 & $-$15.8192 & PL & 2.60 & 7.11 & J180556.4$-$154301 & 271.485 & $-$15.717 & 70 \\
    J1805.9$-$1549 & 271.492 & $-$15.8192 & PL & 2.60 & 7.11 & J180537.9$-$155523 & 271.408 & $-$15.923 & 96 \\
    J1814.2$-$1012 & 273.5617 & $-$10.214 & LP & 3.13 & 9.03 & J181439.4$-$101306 & 273.664 & $-$10.218 & 413 \\
    J1814.2$-$1012 & 273.5617 & $-$10.214 & LP & 3.13 & 9.03 & J181421.6$-$101839 & 273.59 & $-$10.311 & 51 \\
    J1814.2$-$1012 & 273.5617 & $-$10.214 & LP & 3.13 & 9.03 & J181432.1$-$100517 & 273.633 & $-$10.088 & 57 \\
    J1819.9$-$2926 & 274.99 & $-$29.4391 & LP & 2.33 & 4.68 & J182008.7$-$293013 & 275.036 & $-$29.503 & 99 \\
    J1825.2+0715 & 276.3098 & 7.2649 & LP & 2.30 & 1.92 & J182510.0+071533 & 276.291 & 7.259 & 119 \\ 
    J1826.2$-$2830 & 276.57 & $-$28.5161 & LP & 2.62 & 2.81 & J182600.5$-$282114 & 276.502 & $-$28.354 & 43 \\
    J1826.2$-$2830 & 276.57 & $-$28.5161 & LP & 2.62 & 2.81 & J182617.4$-$284143 & 276.573 & $-$28.695 & 44 \\
    J1831.4$-$2909 & 277.864 & $-$29.1552 & LP & 2.38 & 2.45 & J183132.0$-$291327 & 277.883 & $-$29.224 & 41 \\
    J1836.1$-$2656 & 279.039 & $-$26.9484 & LP & 2.54 & 3.10 & J183638.2$-$264543 & 279.159 & $-$26.762 & 159 \\
    J1836.1$-$2656 & 279.039 & $-$26.9484 & LP & 2.54 & 3.10 & J183514.8$-$265425 & 278.811 & $-$26.907 & 625 \\
    J1853.6$-$0620 & 283.424 & $-$6.346 & LP & 2.33 & 4.36 & J185348.4$-$062232 & 283.452 & $-$6.375 & 258 \\
	\hline
\end{tabular}
\end{table*}

\begin{table*}
\centering
\caption{Source names and coordinates of all the gamma-ray sources associated with at least one GLEAM-X: GP radio source, previously targeted by Parkes pulsar search campaigns. The corresponding Parkes project ID is reported in the last column. For each gamma-ray source, the table lists the name (column $1$), the sky coordinates (RA and Dec, columns $2$ and $3$) of the ellipse centre, spectral type (either power-law [PL] or log parabola [LP]), photon index $\Gamma$ (defined as $N(E) \propto (E/E_0)^{-\Gamma}$), and energy flux at $100$\,MeV. For each associated radio source (name in column $7$), the table provides the coordinates (columns $8$ and $9$) and flux density at $200$\,MHz. Sources observed in projects started in 2022 or later are separated by a horizontal line.} \label{tab:observed}
\begin{tabular}{lcccccccccr}
	\hline
    Name & RA & Dec & Spec. & Photon & E$_{100} \times 10^{-12}$ & Name & RA & Dec & S$_{200 MHz}$ & Project\\
    4FGL- & $^{\circ}$ & $^{\circ}$ & Type & Index & erg cm$^{-2}$ s$^{-1}$ & GLEAM-X- & $^{\circ}$ & $^{\circ}$ & mJy & ID\\
	\hline
    J0722.4$-$2650 & 110.614 & $-$26.8441 & LP & 2.20 & 3.73 & J072219.5$-$264936 & 110.581 & $-$26.827 & 31 & P366\\
    %J0722.7$-$2309c & 110.692 & $-$23.1513 & &&& J072328.1$-$230933 & 110.867 & $-$23.1593 & P\\
    J0746.5$-$4113 & 116.644 & $-$41.2256 & PL & 2.08 & 2.18 & J074635.9$-$411453 & 116.65 & $-$41.268 & 310 & P366 \\
    %J0757.9$-$1514 & 119.484 & $-$15.2348 & &&& J075753.5$-$151012 & 119.473 & $-$15.1701 & P366 \\
    J0848.2$-$4527 & 132.072 & $-$45.4656 & PL & 2.13 & 4.93 & J084823.4$-$452432 & 132.098 & $-$45.409 & 314 & P268 \\
    J1202.9$-$5717 & 180.744 & $-$57.296 & LP & 2.51 & 1.71 & J120238.0$-$572607 & 180.658 & $-$57.435 & 237 & P630 \\
    J1202.9$-$5717 & 180.744 & $-$57.296 & LP & 2.51 & 1.71 & J120316.8$-$570959 & 180.82 & $-$57.166 & 19 & P630 \\    
    J1505.1$-$5145 & 226.2877 & $-$51.7552 & LP & 2.41 & 4.03 & J150421.6$-$514759 & 226.09 & $-$51.799 & 179 & P309 \\
    J1505.1$-$5145 & 226.2877 & $-$51.7552 & LP & 2.41 & 4.03 & J150546.4$-$515031 & 226.443 & $-$51.842 & 49 & P309 \\
    J1506.5$-$5708 & 226.636 & $-$57.1474 & LP & 2.29 & 5.29 & J150617.1$-$570748 & 226.571 & $-$57.130 & 35 & P268 \\
    J1517.0$-$4600 & 229.255 & $-$46.0069 & LP & 2.59 & 3.86 & J151735.1$-$460535 & 229.396 & $-$46.093 & 26 & P574 \\
    J1517.0$-$4600 & 229.255 & $-$46.0069 & LP & 2.59 & 3.86 & J151625.7$-$455823 & 229.107 & $-$45.973 & 37 & P574 \\
    J1517.0$-$4600 & 229.255 & $-$46.0069 & LP & 2.59 & 3.86 & J151653.4$-$455902 & 229.222 & $-$45.984 & 179 & P574 \\
    J1517.7$-$4446 & 229.428 & $-$44.7767 & LP & 2.29 & 4.95 & J151750.0$-$444613 & 229.458 & $-$44.770 & 46 & P814 \\  
    J1534.0$-$5232 & 233.5009 & $-$52.5479 & LP & 2.26 & 8.69 & J153354.9$-$523304 & 233.479 & $-$52.551 & 67 & P814 \\
    J1706.2$-$4950 & 256.564 & $-$49.8384 & LP & 2.55 & 3.82 & J170624.6$-$495822 & 256.603 & $-$49.973 & 121 & P268 \\
    J1706.2$-$4950 & 256.564 & $-$49.8384 & LP & 2.55 & 3.82 & J170557.1$-$494618 & 256.488 & $-$49.772 & 84 &  P268 \\
    J1706.2$-$4950 & 256.564 & $-$49.8384 & LP & 2.55 & 3.82 & J170615.1$-$495856 & 256.563 & $-$49.982 & 130 & P268 \\
    J1711.0$-$3002 & 257.77 & $-$30.0487 & LP & 2.19 & 6.43 & J171112.1$-$300329 & 257.800 & $-$30.058 & 137 & P814 \\
    J1716.5$-$5631 & 259.137 & $-$56.5308 & LP & 2.90 & 3.13 & J171621.4$-$562545 & 259.089 & $-$56.429 & 236 & P309 \\
    J1718.5$-$4122 & 259.641 & $-$41.3792 & LP & 2.39 & 4.43 & J171826.8$-$411734 & 259.612 & $-$41.293 & 35 & P268 \\
    J1718.5$-$4122 & 259.641 & $-$41.3792 & LP & 2.39 & 4.43 & J171843.2$-$412637 & 259.68 & $-$41.444 & 22 & P268 \\
    J1729.2$-$2509 & 262.324 & $-$25.1562 & LP & 2.93 & 6.03 & J172903.5$-$250351 & 262.265 & $-$25.064 & 45 & P268 \\
    J1729.2$-$2509 & 262.324 & $-$25.1562 & LP & 2.93 & 6.03 & J172934.7$-$251207 & 262.395 & $-$25.202 & 96 &  P268 \\
    J1729.2$-$2509 & 262.324 & $-$25.1562 & LP & 2.93 & 6.03 & J172946.2$-$250754 & 262.442 & $-$25.132 & 51 & P268 \\
    J1737.5$-$4306 & 264.389 & $-$43.1155 & PL & 2.37 & 2.12 & J173758.9$-$430303 & 264.495 & $-$43.051 & 66 & P630 \\
    J1737.5$-$4306 & 264.389 & $-$43.1155 & PL & 2.37 & 2.12 & J173732.1$-$425903 & 264.384 & $-$42.984 & 142 & P630 \\
    J1745.6$-$3626 & 266.4139 & $-$36.4383 & LP & 2.36 & 5.90 & J174527.5$-$362615 & 266.364 & $-$36.437 & 173 & P050 \\
    J1747.0$-$3505 & 266.771 & $-$35.0989 & LP & 2.64 & 8.32 & J174700.5$-$350553 & 266.752 & $-$35.098 & 42 & P814 \\
    J1752.7$-$3040 & 268.188 & $-$30.6755 & LP & 2.52 & 6.83 & J175259.5$-$303836 & 268.248 & $-$30.643 & 53 & P630 \\
    J1752.7$-$3040 & 268.188 & $-$30.6755 & LP & 2.52 & 6.83 & J175304.2$-$304213 & 268.267 & $-$30.704 & 42 & P630 \\
    J1755.3$-$3937 & 268.828 & $-$39.6222 & LP & 2.56 & 2.68 & J175445.2$-$394031 & 268.688 & $-$39.675 & 240 & P050 \\
    J1755.3$-$3937 & 268.828 & $-$39.6222 & LP & 2.56 & 2.68 & J175452.5$-$394251 & 268.719 & $-$39.714 & 107 & P050 \\
    J1759.6$-$1850 & 269.924 & $-$18.8409 & PL & 2.55 & 7.13 & J175907.2$-$184922 & 269.78 & $-$18.823 & 120 & P268 \\
    J1759.6$-$1850 & 269.924 & $-$18.8409 & PL & 2.55 & 7.13 & J175932.1$-$185705 & 269.884 & $-$18.951 & 60 & P268 \\
    J1801.1$-$3740 & 270.277 & $-$37.6769 & LP & 2.63 & 3.06 & J180059.4$-$374221 & 270.248 & $-$37.706 & 741 & P050 \\
    J1802.4$-$3041 & 270.6147 & $-$30.6992 & LP & 1.81 & 6.30 & J180224.6$-$304306 & 270.602 & $-$30.718 & 110 & P268 \\
    J1803.5$-$1639 & 270.887 & $-$16.6559 & LP & 2.55 & 5.39 & J180314.1$-$164158 & 270.809 & $-$16.699 & 124 & P268 \\
    J1803.5$-$1639 & 270.887 & $-$16.6559 & LP & 2.55 & 5.39 & J180345.8$-$164527 & 270.941 & $-$16.757 & 45 & P268 \\
    J1808.2$-$1055 & 272.0582 & $-$10.9222 & LP & 3.02 & 9.77 & J180821.5$-$105051 & 272.089 & $-$10.847 & 45 & P268 \\
    J1817.9$-$3334 & 274.4799 & $-$33.5727 & LP & 2.25 & 4.83 & J181810.3$-$333501 & 274.543 & $-$33.584 & 444 & P814 \\
    J1834.3$+$0613 & 278.5757 & 6.2256 & LP & 2.16 & 3.30 & J183411.6$+$061747 & 278.548 & 6.296 & 63 & P858 \\
    J1846.0$+$0507 & 281.5075 & 5.1278 & LP & 1.93 & 4.06 & J184558.7$+$050455 & 281.495 & 5.082 & 34 & P630 \\
    J1851.9$-$1522 & 282.9869 & $-$15.3726 & LP & 2.66 & 2.58 & J185143.1$-$151605 & 282.929 & $-$15.268 & 68 & P309 \\
    J1851.9$-$1522 & 282.9869 & $-$15.3726 & LP & 2.66 & 2.58 & J185145.9$-$153234 & 282.941 & $-$15.543 & 40 & P309 \\ 
    J1925.1$+$0547 & 291.2908 & 5.784 & LP & 2.74 & 4.41 & J192502.2$+$055410 & 291.259 & 5.903 & 60 & P268 \\
    J1925.1$+$0547 & 291.2908 & 5.784 & LP & 2.74 & 4.41 & J192502.3$+$055547 & 291.259 & 5.929 & 57 & P268 \\
	\hline
    J0754.9$-$3953 & 118.741 & $-$39.8951 & LP & 1.98 & 3.19 & J075452.5$-$395316 & 118.719 & $-$39.888 &  58 & P1194 \\
    J1208.0$-$6900 & 182.0204 & $-$69.0034 & LP & 2.28 & 5.73 & J120750.0$-$690007 & 181.958 & $-$69.002 & 45 & P1054 \\
    J1517.9$-$5233 & 229.4886 & $-$52.5548 & LP & 1.91 & 8.75 & J151805.9$-$523343 & 229.524 & $-$52.562 & 21 & P1211 \\
    J1517.9$-$5233 & 229.4886 & $-$52.5548 & LP & 1.91 & 8.75 & J151754.2$-$523230 & 229.476 & $-$52.542 & 14 & P1211 \\
    J1815.8$-$1416 & 273.971 & $-$14.275 & LP & 2.55 & 14.5 & J181556.9$-$141636 & 273.987 & $-$14.277 & 84 & P1348 \\
    \hline
\end{tabular}
\end{table*}

%Each of these will be inspected in detail in follow-up analysis and potential pulsation searches.

\section{Results and discussion} \label{sec:results}
In the following sections, we are going to discuss the results of the cross-matches performed in this work. We first focus on the associations within radio sources and known pulsars in the ATNF catalogue. We highlight the improvements that low-frequency measurements have brought to the spectra of the known population, and constrain some parameter distributions to show how pulsars with specific characteristics are likely to be detected with our approach. Secondly, we will focus on the identification of a probable pulsar candidate through the cross-match with the unidentified catalogue of gamma-ray sources. We will highlight the distribution of the parameters used for the filtering and the properties of the radio sources that have been found to be localised within the Fermi ellipses.

\subsection{Known pulsars} \label{sec:discussion}
We analysed the spectral behaviour of the $193$ known pulsars detected in the GLEAM-X: GP catalogue using \textit{pulsar$\_$spectra}, as described in \sect~\ref{sec:software}. Since the development of this tool, $8$ of the detected pulsars have been newly added to the ATNF catalogue. As a result, these pulsars lack the prior literature necessary for informed spectral fitting. In these cases, we performed fits using the flux density values reported directly in the ATNF catalogue if present. 

\begin{figure}
    \centering
    \subfloat[][\emph{J1534$-$46}]
    {\includegraphics[scale=0.7]{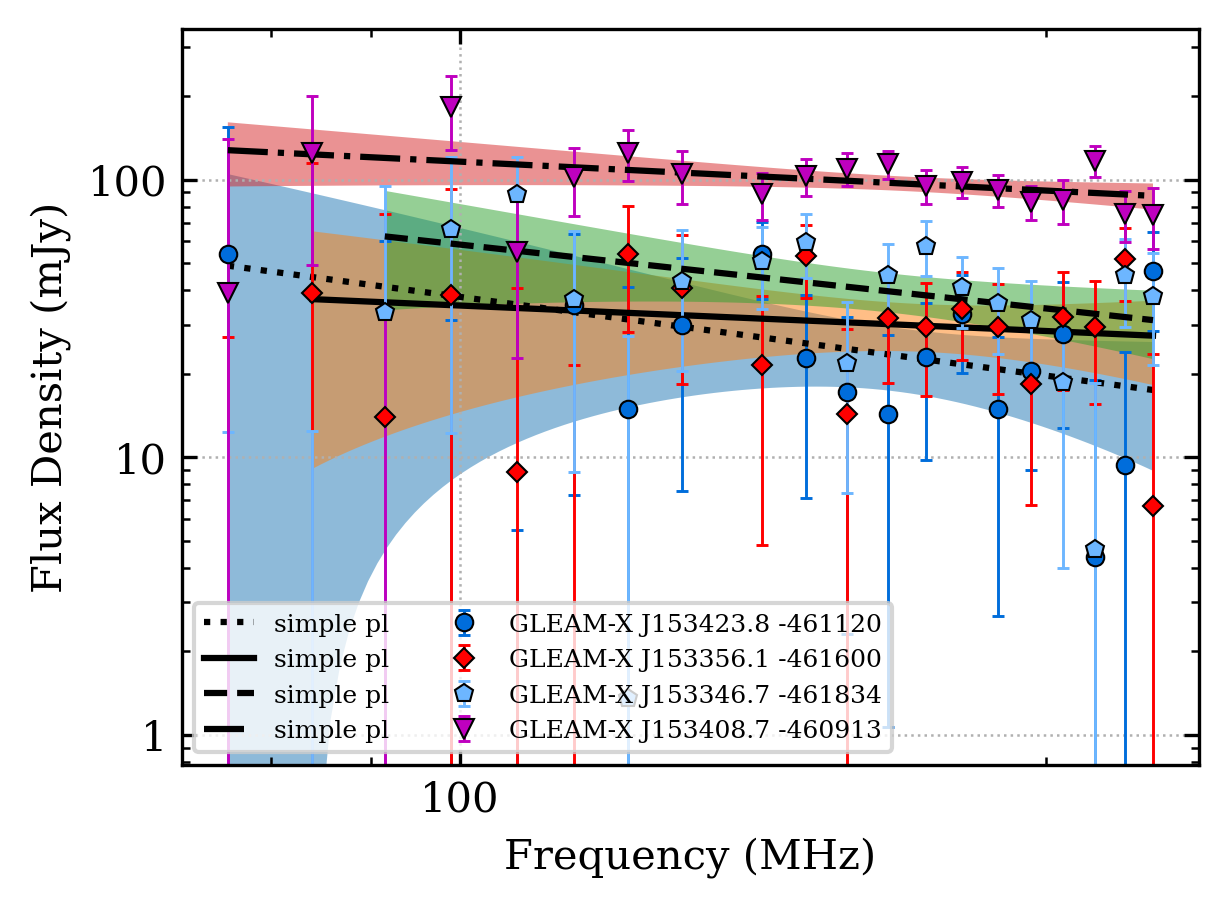}} \\
    \subfloat[][\emph{J1708$-$52}]
    {\includegraphics[scale=0.7]{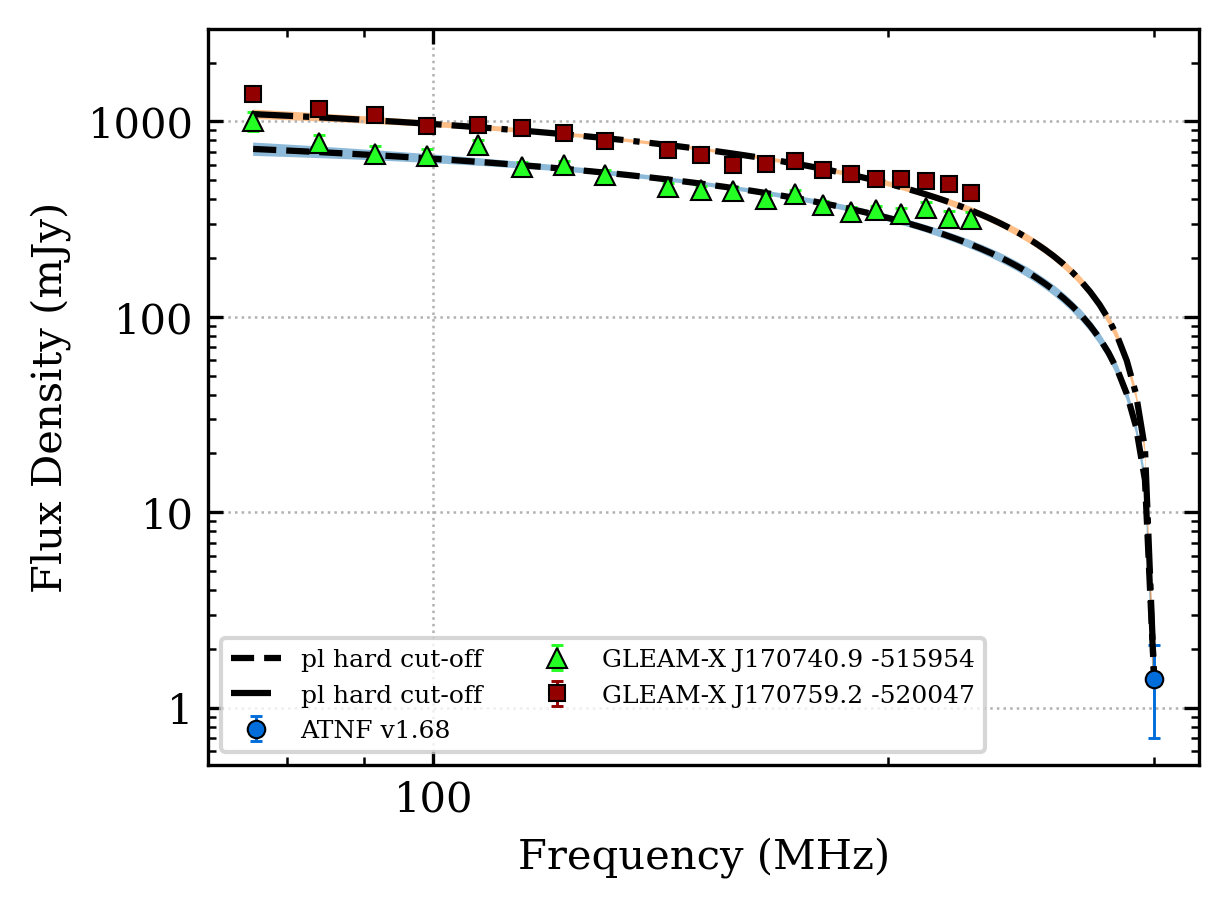}} \\

    \caption{Spectral distributions of the two pulsars with multiple associations. The lack of high-frequency data points makes the matched sources equally believable. All the sources in the GLEAM-X: GP catalogue are shown in a different colour to make a clear distinction.}\label{fig:seds-1}
\end{figure}

J1708$-$52 and J1534$-$46 had 2--4 potential matches in the GLEAM-X: GP catalogue due to large positional uncertainties or high source density in the sky region. For these, we performed spectral fitting for each possible low-frequency counterpart to help estimate the most likely counterpart based on spectral information, and the SEDs are shown in \fig~\ref{fig:seds-1}. The associated radio sources of J1708$-$52 are quite bright ($\simeq$~Jy) in opposition to J1534$-$46, for which the flux of the matches is low (below $\approx 100$\,mJy), which suggests they have probably been missed before because of their faintness and confusion with the surrounding environment. The SEDs can help assess the likelihood of chance alignments by inspecting the quality of the fit across all available data points. In the cases where no ATNF data points are present, or only a single data point is available, it is not possible to claim with confidence whether a particular source is the genuine counterpart. The remaining $193$ pulsars each had a single, unambiguous match. Their positions are well constrained by the $3$~arcseconds astrometry of the GLEAM-X: GP catalogue, allowing for confident identification. The corresponding spectral plots for all pulsars are available for download online through the MNRAS data store. 
%~\footnote{https://www.dropbox.com/scl/fo/d1rk0yindzvthxp728wka/AHPJZzMHpC 73Zp8k4P5JwP0?rlkey=bjwllmma02xshzbiqbf5wreza\&st =xv2iunip\&dl=0}.

\begin{figure}
    \centering
    \subfloat[][]{\includegraphics[scale=0.3]{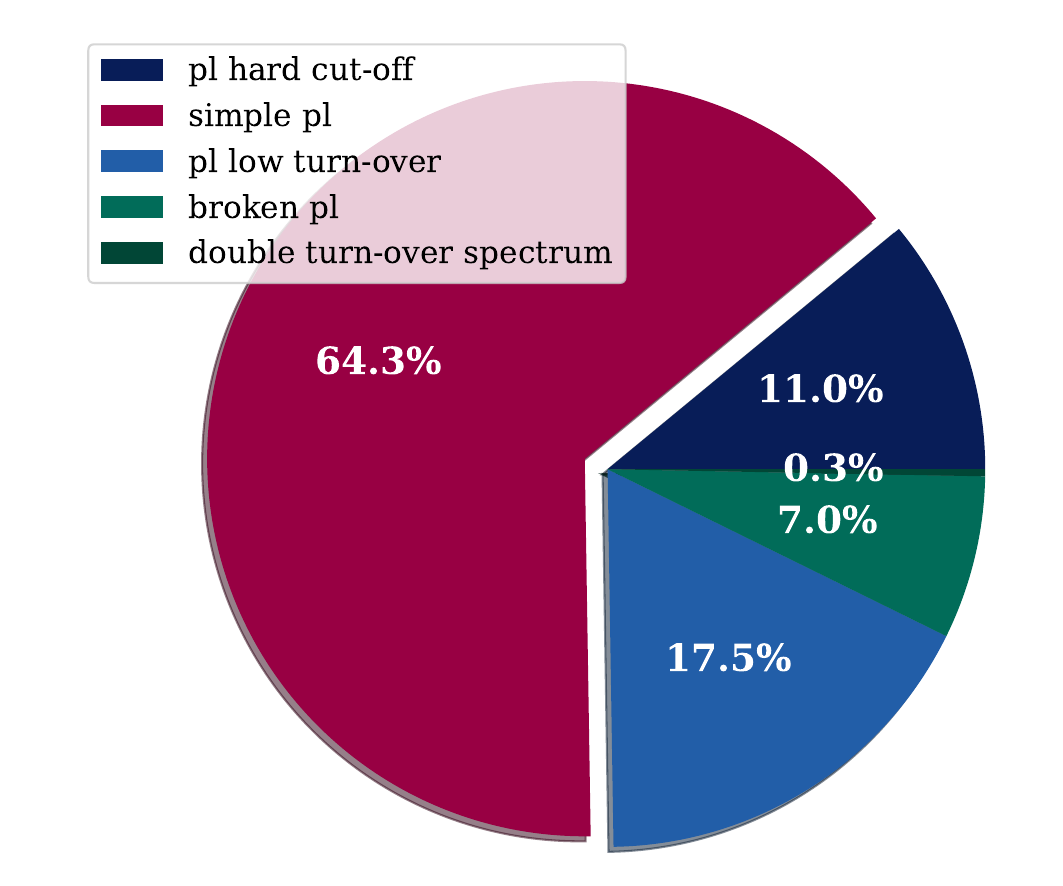}} \\
    \subfloat[][]{\includegraphics[scale=0.3]{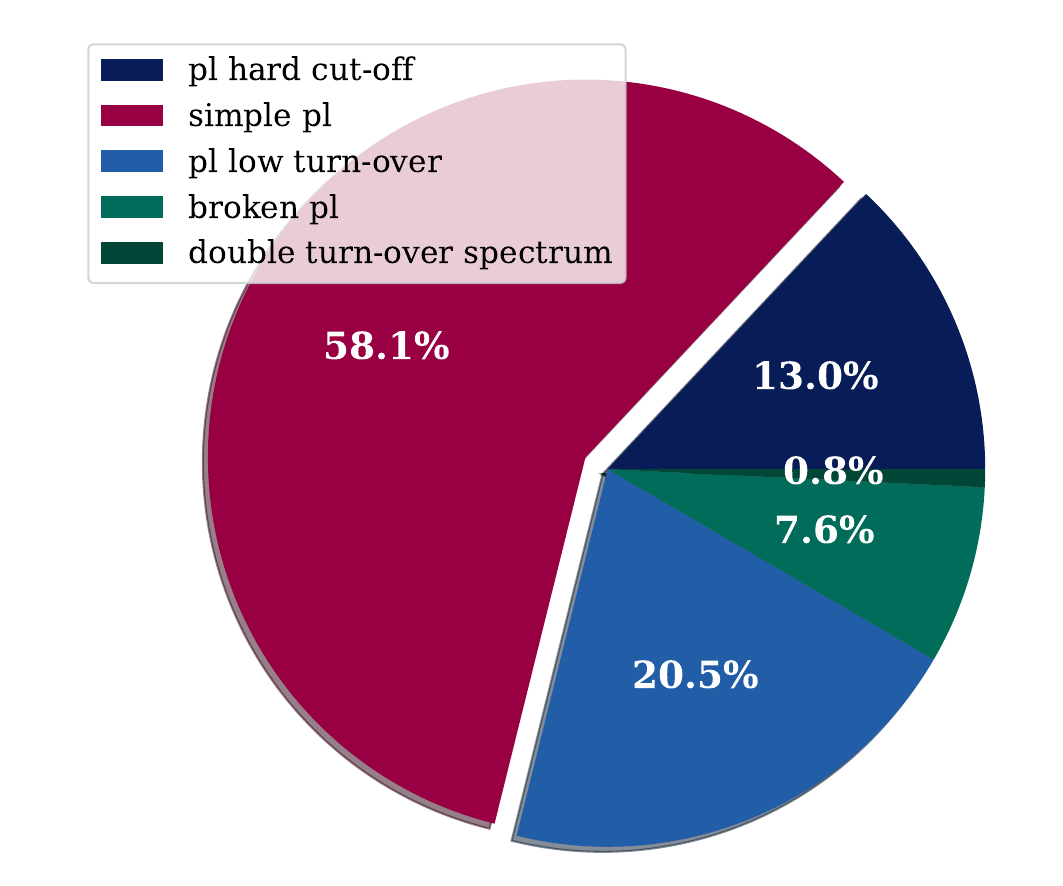}} \\

    \subfloat[][]{\includegraphics[scale=0.3]{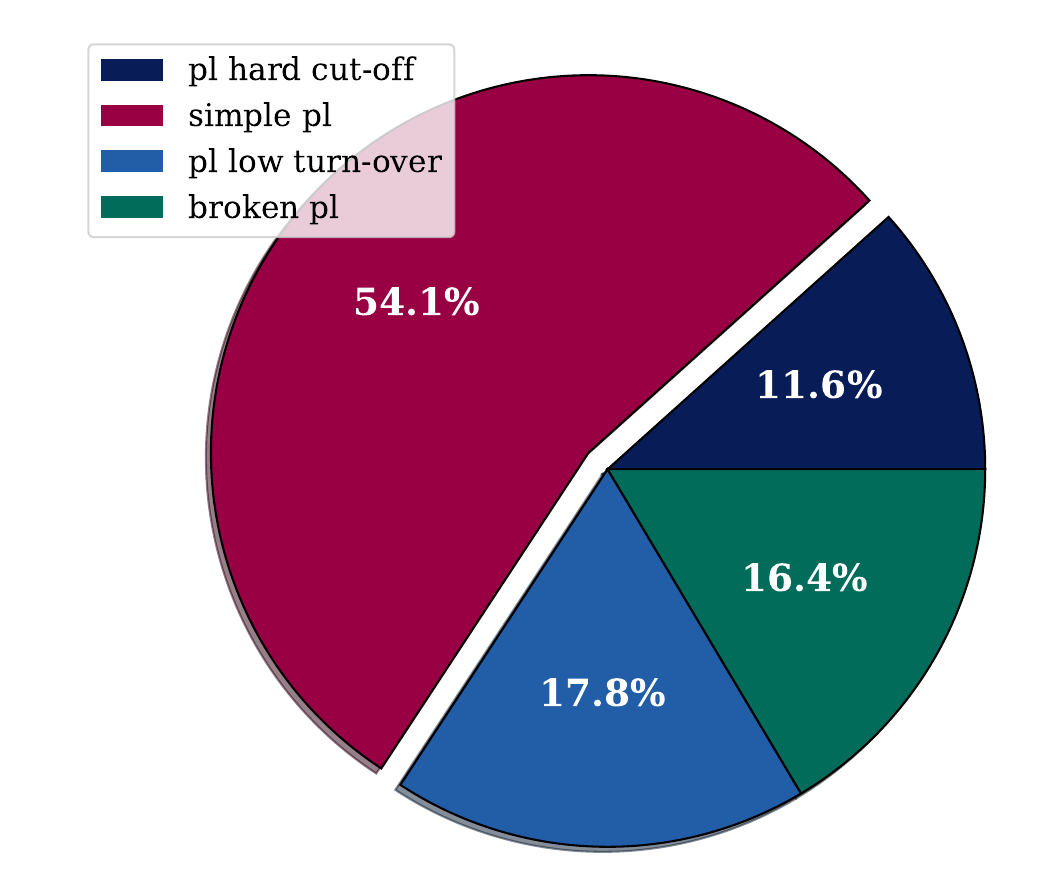}} \\
    \subfloat[][]{\includegraphics[scale=0.3]{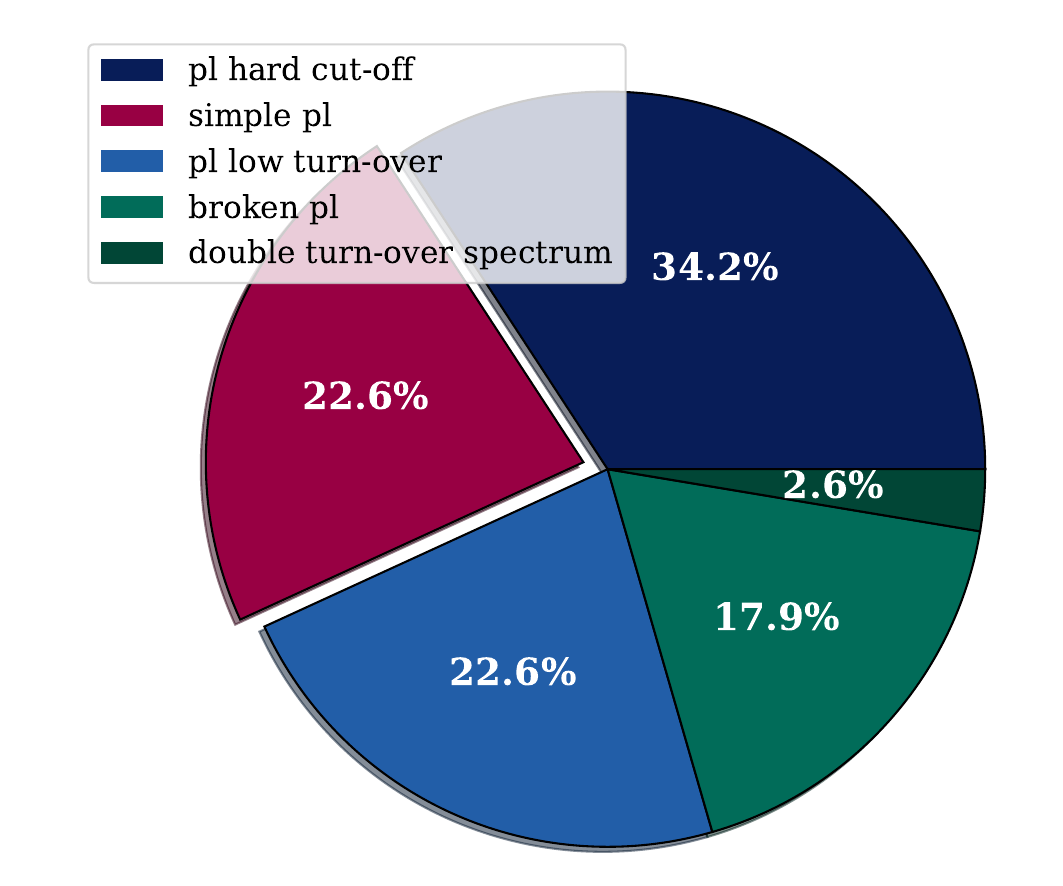}} \\
    \caption{Percentage of pulsars fitted by each model across different datasets. The top two charts display results for all known pulsars from the ATNF catalogue, without (a) and with (b) low-frequency data. The bottom two panels show only sources matched with the GLEAM-X: GP catalogue, also without (c) and with (d) low-frequency data.} \label{fig:models}
\end{figure}

The majority of the matched pulsars were best modelled by a power law exhibiting either a low-frequency turnover or a high-frequency cutoff, as shown in the top pie chart of \fig~\ref{fig:models}. Only $23 \%$ of pulsars are well fitted by a simple power law with an average spectral index of $-1.7 \pm 0.7$, highlighting the complexity of their emission mechanisms and/or the significant role of absorption effects in this frequency regime. For comparison, we extended the spectral fitting to include all the remaining pulsars in the ATNF catalogue. Out of the $\approx 3,600$ sources, only around 850 were fitted by one of the spectral models. This limited success is primarily caused by the sparse data coverage for most pulsars, with many having only three or fewer measured flux density points. As illustrated in the bottom pie chart of \fig~\ref{fig:models}, most of them are best described by a simple power-law model, with an average spectral index of $-1.6 \pm 0.9$. This result highlights the value of low-frequency observations in revealing more detailed spectral features and improving our understanding of pulsar emission.

Of particular note, $106$ pulsars were detected below $400$\,MHz for the first time, including $36$ first detected below $300$\,MHz. These low-frequency detections provide valuable new constraints on pulsar spectra. Finally, we report flux density measurements for one pulsar (J1901$-$0125) for the first time, previously missed due to its steep spectral index of $-2.8 \pm 0.1$. 
%We reported the flux density measurements of this source in Table~\ref{tab:fluxes}, including each of the $20$~sub-bands of GLEAM-X: GP. 

\begin{figure*}
\centering
\includegraphics[width=1.0\linewidth]{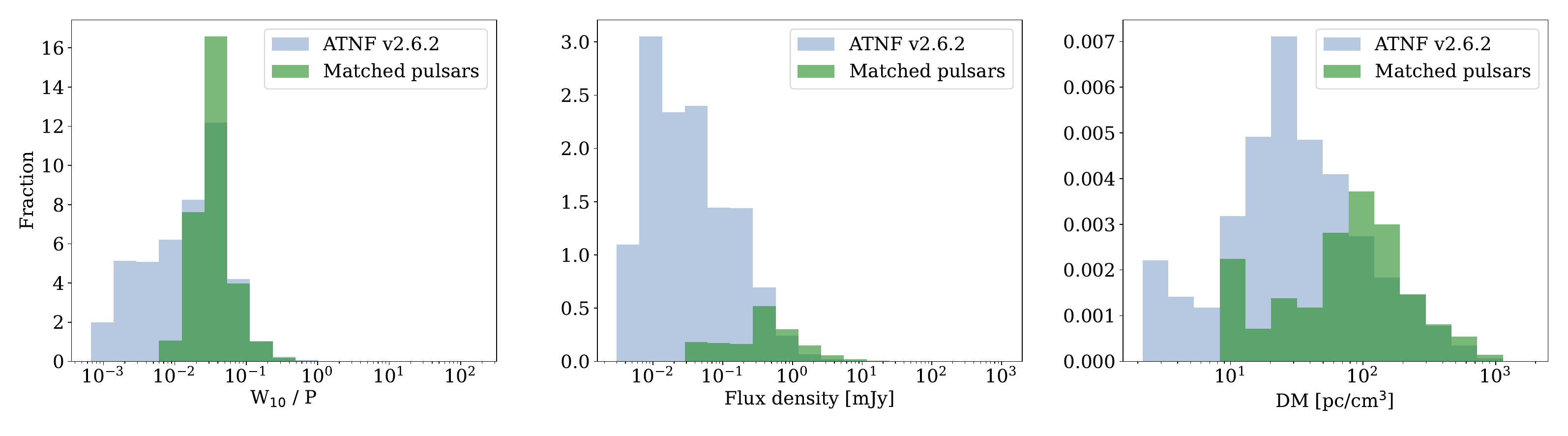}
\caption{Distribution of the fractional pulse width (left), flux density (middle), and DM (right) for the matched pulsars, shown in comparison with the known pulsar population from the ATNF catalogue (version 2.6.2). All histograms are represented on a logarithmic scale.} \label{fig:dm-fl}
\end{figure*}

We also analysed the characteristic properties of the pulsars detected in this work to identify potential indicators suggesting a preference for pulsar searches in the image domain. First, we considered the fractional pulse width, defined as the pulse width at $10\%$ of the peak ($W_{10}$) divided by the period. The left panel of \fig~\ref{fig:dm-fl} shows that our detections tend to group at higher fractional widths compared to the broad known population. Next, as shown in the middle panel of \fig~\ref{fig:dm-fl}, we primarily detected the brightest pulsars with a clear skew toward higher flux densities at $1400$\,MHz. This trend reflects the sensitivity limitations of image-domain techniques, which naturally favour sources with strong continuum emission. Finally, we observed a preference for pulsars with higher DMs as illustrated in the right panel of \fig~\ref{fig:dm-fl}. As such, image methods may offer a complementary advantage in probing more distant or heavily scattered pulsars that are otherwise challenging to detect through standard periodicity searches.  

\subsection{High energy candidates}
\begin{figure}
\centering
\includegraphics[width=1.0\linewidth]{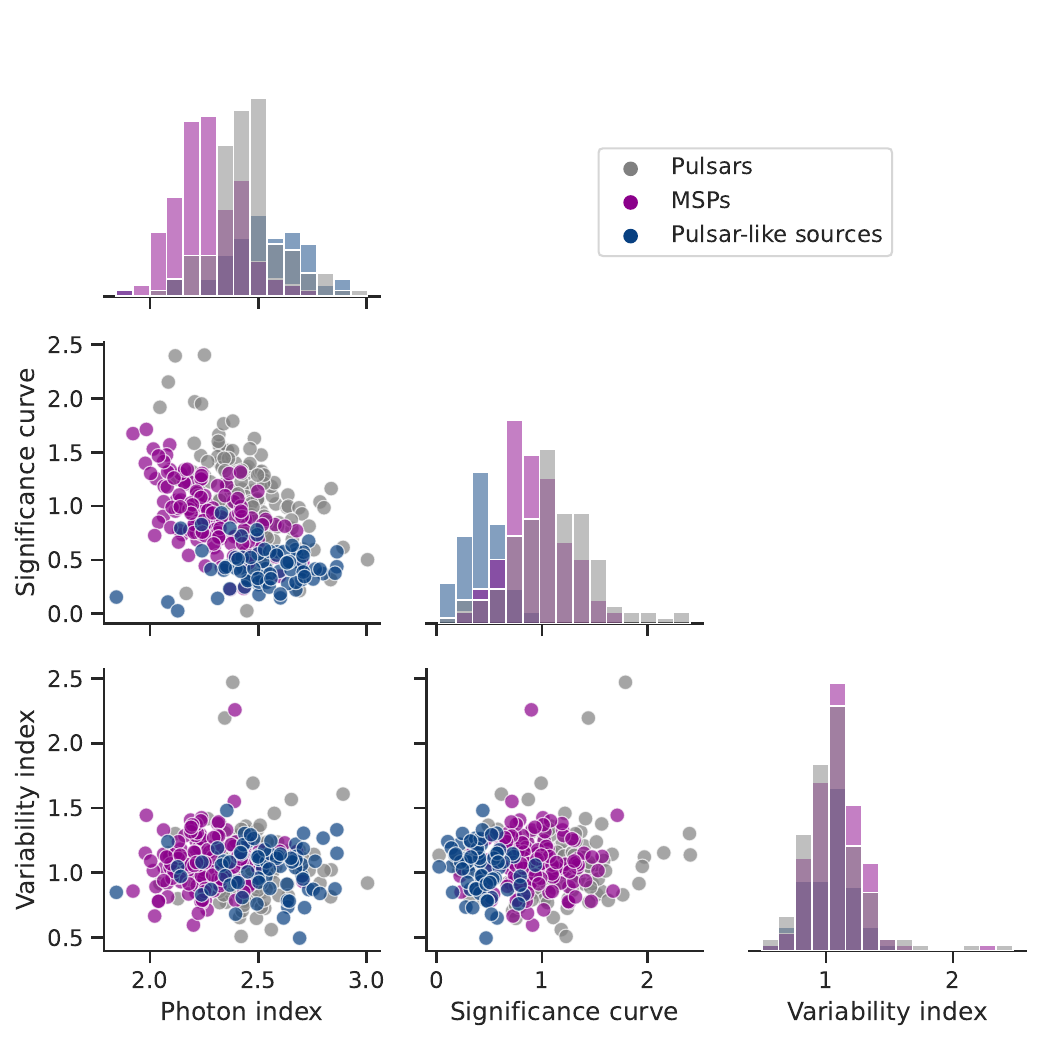}
\caption{Corner plot illustrating the distribution of key parameters (photon index, significance curve, and variability index) used in this work to filter the sample of pulsar-like sources (blue), shown in comparison with known pulsars (grey) and MSPs (magenta). Diagonal panels display histograms for each parameter, while the off-diagonal panels show scatter plots illustrating the correlations between parameter pairs. The variability index and significance curve are reported in log values.} \label{fig:params}
\end{figure}

The cross-matching procedure identified $73$ unique gamma-ray sources as likely pulsar candidates, with a total of $106$ associated radio counterparts in the GLEAM-X: GP catalogue. We compare this candidate sample with the known gamma-ray pulsars listed in the 4FGL-DR4 catalogue. \fig~\ref{fig:params} presents scatter plots of the three key parameters (significance curve, variability index, and photon index) used to filter the 4FGL catalogue, showing the distributions for known pulsars, known MSPs, and the pulsar-like sources identified in this work. The candidate sources occupy an outlier region of the parameter space relative to the known populations, primarily due to the significance curve parameter, which quantifies the significance of spectral curvature for every source modelled with a log-parabola. While low significance values may indicate AGN-like characteristics, such sources should not be excluded when considered in the context of other diagnostic parameters. Notably, some pulsars may exhibit spectra with cutoffs beyond the LAT energy range, resulting in spectra that appear as straight power laws within the observed band. Our methodology is then effective in identifying new pulsar candidates without imposing strong conditions on high-energy spectral properties, such as requiring specific spectral models or energy thresholds, thus enabling the inclusion of sources that might otherwise be overlooked.

Furthermore, we did not impose restrictions based on the size of the gamma-ray error ellipses. The only cases we excluded were those with several radio candidates within the positional uncertainty region. However, for the retained sources, we provide precise radio positions, which are essential for follow-up pulsation searches. Given that the maximum semi-major axes of the gamma-ray error ellipses in our sample are up to $0.5$~degrees, accurate localisation is crucial; even with the frequency-dependent wide field of the new cryogenically cooled phased array feed (CryoPAF: $\approx 1.5$ sq. degrees) for the Parkes telescope, conducting a blind pulsation search over such large areas would be time- and resource-intensive.

All associated radio sources have been fitted using either a simple power-law or a curved power-law model, with the best-fit parameters reported in the GLEAM-X: GP catalogue. Notably, GLEAM-X~J075306.2$-$465803 is the only source in our sample that exhibits a curved radio spectrum. We examined the radio spectral indices of each source and found that all values fall within the range reported by \cite{Bates2013} ($-1.41 \pm 0.96$), with the exception of eight sources. Nevertheless, these outliers remain strong candidates, as demonstrated in Section~\ref{sec:discussion}: only a small fraction (in our case $27\%$) of known pulsars follow a simple power-law radio spectrum, while the majority show spectral turnover at low frequencies or exhibit a hard cutoff. In addition, none of the radio candidates in our sample can be classified as radio-quiet. When extrapolating their flux densities to $1400$\,MHz using the spectral indices reported in the GLEAM-X: GP catalogue, all sources yield flux densities exceeding $30\mu$Jy. This suggests that, if confirmed as pulsars, they would be detectable in radio gating searches.

It is also important to note that we did not apply any spatial filtering during the selection process. While pulsars, and particularly MSPs, are statistically more commonly found at high Galactic latitudes because they are less affected by dispersion and scattering, $25$ out of our $73$~pulsar-like sources are located at low latitudes ($|b| < 4^{\circ}$). 

%We can make a classification of the likelihood of being a good candidate by looking at the high-energy spectra information. These properties include whether they have been fitted by a high-energy cutoff spectrum with minimal emission above 10 GeV, if they have a little excess, and if they show peak significance between 1 and 5 GeV.

\section{Conclusions}\label{sec:conclusion}
We present spectral information for $193$~known pulsars detected in the GLEAM-X: GP data release, including $106$ detected for the first time below $400$\,MHz. These low-frequency measurements are essential in characterising pulsar spectra, as many sources exhibit spectral turnovers or hard cutoffs that weren't recognised in previous statistics. 

We also present a list of $106$~compact radio sources located within Fermi LAT error ellipses, selected based on gamma-ray spectral and variability properties as promising pulsar candidates. Future follow-up observations, particularly targeted pulsation searches, are essential to confirm the nature of these sources.

\section*{Acknowledgements}
N.H.-W. is the recipient of an Australian Research Council Future Fellowship (project number FT190100231).
This research has made use of the SIMBAD database, operated at CDS, Strasbourg, France.
This paper includes archived data obtained through the Parkes Pulsar Data archive on the CSIRO Data Access Portal (http://data.csiro.au).
This scientific work uses data obtained from Inyarrimanha Ilgari Bundara, the CSIRO Murchison Radio-astronomy Observatory. We acknowledge the Wajarri Yamaji People as the Traditional Owners and Native Title Holders of the observatory site. Support for the operation of the MWA is provided by the Australian Government (NCRIS), under a contract to Curtin University administered by Astronomy Australia Limited. We acknowledge the Pawsey Supercomputing Centre which is supported by the Western Australian and Australian Governments.

\section*{Data Availability}
The pulsar–radio source associations are listed in a catalogue and made available in machine-readable format via MNRAS. The catalogue includes pulsar names and associated GLEAM-X source names, source positions in both catalogues, and flux densities in all 20 sub-bands covered by the GLEAM-X survey. All radio source data were derived from the GLEAM-X survey, which is available on the AAO Data Central and Vizier. The corresponding SEDs for all pulsars are available for download online through the MNRAS data store. 
%at \url{https://www.dropbox.com/scl/fo/d1rk0yindzvthxp728wka/AHPJZzMHpC 73Zp8k4P5JwP0?rlkey=bjwllmma02xshzbiqbf5wreza\&st =xv2iunip\&dl=0}.

\bibliographystyle{mnras}
\bibliography{biblio}

% Don't change these lines
\bsp
\label{lastpage}
\end{document}